\documentclass[reprint,
showpacs,
amsmath,amssymb,prd]{revtex4-1}

\usepackage{graphicx}
\usepackage{dcolumn}
\usepackage{bm}
\usepackage{esdiff}
\usepackage{hyperref}
\usepackage[mathlines]{lineno}
\usepackage{placeins}
\graphicspath{{./}{figures/}}
\newcommand{\ts}{\textsuperscript}
\begin{document}

\title{On analytic solutions of wave equations in regular coordinate systems on Schwarzschild background}

\author{Dennis Philipp}
\email{dennis.philipp@zarm.uni-bremen.de}

\author{Volker Perlick}
\email{volker.perlick@zarm.uni-bremen.de}

\affiliation{ZARM, University of Bremen, 28359 Bremen, Germany}
\date{\today}

\begin{abstract}
The propagation of (massless) scalar, electromagnetic and gravitational waves on fixed Schwarzschild background spacetime is described by the general time-dependent Regge-Wheeler equation. We transform this wave equation to usual Schwarzschild, Eddington-Finkelstein, Painlev\'{e}-Gullstrand and Kruskal-Szekeres coordinates. In the first three cases, but not in the last one, it is possible to separate a harmonic time-dependence. Then the resulting radial equations belong to the class of confluent Heun equations, i.e., we can identify one irregular and two regular singularities. Using the generalized Riemann scheme we collect properties of all the singular points and construct analytic (local) solutions in terms of the standard confluent Heun function HeunC, Frobenius and asymptotic Thom\'{e} series.
We study the Eddington-Finkelstein case in detail and obtain a solution that is regular at the black hole horizon. This solution satisfies causal boundary conditions, i.e., it describes purely ingoing radiation at $r=2M$.
To construct solutions on the entire open interval $r \, \in \; ]0,\infty[ \,$, we give an analytic continuation of local solutions around the horizon. Black hole scattering and quasi-normal modes are briefly considered as possible applications and we use semi-analytically calculated graybody factors together with the Damour-Ruffini method to reconstruct the power spectrum of Hawking radiation emitted by the black hole.
\end{abstract}

\maketitle


\section{\label{sec:intro} Introduction}
While many features of black hole spacetimes can be investigated in terms of the motion of (massive and massless) particles, wave propagation opens a perspective on additional phenomena such as interference effects, scattering of radiation at a black hole, its quasi-normal modes and black hole evaporation. Scattering and interference patterns have proven to be very important tools in other branches of physics and yield observational properties of different physical systems. Therefore, it is of great relevance to consider black-hole scattering as well and to describe the dynamics of fields in the presence of a black hole for different boundary and initial conditions.

The central equation that describes the propagation of massless waves on Schwarzschild background spacetime is the general (time-dependent) Regge-Wheeler equation which is a  wave equation written in the usual Schwarzschild time and the so-called  tortoise coordinate. Regge and Wheeler found this equation for spin $s=2$  by investigating the stability of a Schwarzschild black hole under gravitational perturbations in a linearized perturbation theory \cite{ReggeWheeler1957}. To study electromagnetic wave propagation one can solve Maxwell's equations on flat spacetime but consider a medium with specified constitutive relations \cite{Plebanski1960,Mashhoon1973}. Another possibility to derive the Regge-Wheeler equation for $s=1$ is to use the so-called Debye equation, because the source-free Maxwell equations reduce to this single scalar equation on a Schwarzschild background \cite{MoPapas1971, Stephani1974}. 
For scalar perturbations and their propagation the Klein-Gordon equation has to be employed and the Regge-Wheeler equation for $s=0$ follows from a straight-forward calculation \cite{FrolovNovikovBook}.
For each of the three different types of perturbations the respective Regge-Wheeler-type equation follows after an expansion of the fields into (tensorial, vectorial and scalar) spherical harmonics.
Assuming additionally a harmonic time dependence, i.e., Fourier expanding the solution of this wave equation, leads to a radial equation that is known as the stationary Regge-Wheeler equation. This radial equation is a Schr\"odinger-type differential equation with a spin-dependent potential barrier. Due to the form of this potential no exact solution is known as long as the tortoise coordinate is kept as the radial variable. 

Changing the coordinate system and reconsidering the previous wave equations gives rise to new radial equations that admit analytic solutions in terms of series of special functions such as hypergeometric or Coulomb wave functions, see e.g. Ref. \cite{Leaver1986}. Another very insightful method is to use the Heun functions \cite{Fiziev2006, Fiziev2007} and the theoretical framework of singularity analysis to construct analytic (local) solutions to the radial equations. We will follow this strategy to a great extent in this article.

We start with some necessary preparations in Sec. \ref{sec:prep} where we introduce the confluent Heun equation and important notations concerning singularities of differential equations and local solutions around them.
Furthermore, we consider on a Schwarzschild background a wave equation in the tortoise coordinate with an arbitrary potential term. This equation is transformed to usual Schwarzschild coordinates. By considering a general coordinate transformation that keeps the angles unchanged we proceed to Eddington-Finkelstein, Painlev\'{e}-Gullstrand and Kruskal-Szekeres coordinates. 
In the last part of this section we sketch how to derive the respective wave equations for the three different spin values. We combine the results to obtain the general Regge-Wheeler equation that has the form of the previously discussed wave equation in the tortoise coordinate and can, thus, be transformed to the respective wave equations in the other coordinate systems.

In Sec. \ref{sec:radialEqn1} we study the radial equation in the tortoise coordinate, which is the stationary Regge-Wheeler equation. A lot about the phenomenology can be learned from this simplest form of the radial equation. We describe asymptotic solutions that fulfill the causal boundary condition at the horizon, the so-called IN-mode for scattering and the QNM-asymptotics. Here and in the following,  QNM stands for quasi-normal modes. Since there is no known exact analytic solution of the stationary Regge-Wheeler equation beyond the asymptotic behavior,  approximations or numerical methods have to be used to obtain the full solutions of this equation.
Afterwards, we derive the radial equation in the usual Schwarzschild coordinate, which was already considered e.g. by Leaver \cite{Leaver1985, Leaver1986} and Fiziev \cite{Fiziev2006, Fiziev2007}. We briefly summarize their results and give properties as well as local solutions of this radial equation. All these solutions have only finite and non-sufficient domains of convergence and there is no regular solution at the black hole horizon that allows to penetrate this radius.

In Sec. \ref{sec:radialEqn2} we derive and discuss the radial equation in Eddington-Finkelstein coordinates in full detail, and we examine its properties and local solutions in terms of Frobenius and Thom\'{e} series and using the confluent Heun function $\mathrm{HeunC}$. 
As far as we know, there is no previous work concerned with the analysis of the Eddington-Finkelstein radial equation to find analytic solutions of the perturbation equations on Schwarzschild background,
apart from a short remark on their asymptotic behavior at the horizon in Ref. \cite{DamourRuffini1976}. To overcome the fact that the region where the local solutions converge is bounded, we show how an analytic continuation of the local Frobenius solutions onto the entire interval $r \in \, ]0,\infty[ \,$ can 
be performed. In this way we construct an exact solution on this interval that is regular at the horizon and satisfies the causal boundary condition.

In Sec. \ref{sec:applications} we examine some applications of our analytically continued solutions. We consider black hole scattering and briefly sketch how to calculate QNM-frequencies for the different perturbations. At the end of this section we use the method of Damour and Ruffini \cite{DamourRuffini1976} and our semi-analytically calculated graybody factors to reobtain the power spectrum of the Hawking radiation emitted from a Schwarzschild black hole.
\section{\label{sec:prep} Preparation}
\subsection{The confluent Heun equation and its properties}
We consider an ordinary, second order, linear and homogeneous differential equation of the form
\begin{align}
\left[ \diff[2]{}{z} + p_1(z) \diff{}{z} + p_0(z) \right] y(z) = 0 \label{eq:ODE}
\end{align}
with rational coefficient functions $p_i(z)$, where $z$ shall be an element of the Riemann sphere, i.e., of the extended complex plane including $z=\infty$. 
Following \cite{BenderOrszag1978}, a point $z=z_0$ 
\begin{align*}
\text{is }
\begin{cases}
\text{an ordinary point  if } \forall i \;  p_i(z) \text{ is analytic at } z=z_0 \, \\
\text{a singularity otherwise.} 
\end{cases}
\end{align*}
A singularity at the point $z=z_j$ 
\begin{align*}
\text{is }
\begin{cases}
\text{regular if } \forall i \; (z-z_j)^{2-i} \, p_i(z) \text{ is analytic at } z=z_j \\
\text{irregular otherwise.}
\end{cases}
\end{align*}
If a differential equation possesses only regular singularities it is referred to as being Fuchsian, otherwise it is called a confluent equation. The singly confluent Heun equation (CHE) belongs to the remarkably large class of Heun's differential equations. All members of this class can be derived from a single second order Fuchsian differential equation of the form \eqref{eq:ODE} with four singularities, the general Heun equation (GHE), see e.g. \cite{Heun1888, RonveauxBook,  SlavyanovLayBook, MapleHeun}. There are four different confluent cases which are obtained through (consecutive) confluence processes of regular singularities.
In the case of the CHE, two regular singularities of the GHE, one of them located at infinity, merge and form an irregular singularity. One standard form of this CHE is given by \eqref{eq:ODE}, where the coefficient functions are 
\begin{subequations}
\label{eq:CHE}
\begin{align}
p_1(z) &= \dfrac{\gamma}{z} + \dfrac{\delta}{z-1} - \beta \, , \\
p_0(z) &= - \left( \dfrac{\alpha \beta -q}{z-1} + \dfrac{q}{z} \right) \, ,
\end{align}
\end{subequations} 
and regular singularities are located at $z=0,1$. By considering the transformation $\zeta = 1/z$ we find an irregular singularity at $\zeta = 0$, i.e., $z = \infty$.
The three singularities can be further characterized by their singular rank (s-rank).  The s-rank of a regular singularity is by definition always equal to one, so we only have to calculate the rank of the irregular singularity. For the equation in the form \eqref{eq:CHE} this s-rank is found to be two, where we follow the convention in Ref. \cite{SlavyanovLayBook}. The generalized Riemann scheme (GRS), in the form proposed by Slavyanov and Lay \cite{SlavyanovLayBook}, is a very useful tool in the analysis of singular differential equations. It contains
\begin{itemize}
\setlength{\parskip}{1pt}
\item[-] s-ranks and locations of all singularities (1\ts{st} and 2\ts{nd} row), 
\item[-] indicial (characteristic) exponents of the regular singularities (3\ts{rd} and 4\ts{th} row) and
\item[-] characteristic exponents of second kind for the irregular singularity (3\ts{rd} row and following).
\end{itemize}
The corresponding GRS for the confluent Heun equation in the form \eqref{eq:CHE} is
\begin{align}
\begin{pmatrix}
1 & 1 & 2 &  \\ 
0 & 1 & \infty & ;z \\ 
0 & 0 & \alpha & ;q \\ 
1-\gamma & 1-\delta & \gamma+\delta-\alpha &  \\ 
 &  & 0 &  \\ 
 &  & \beta & 
\end{pmatrix} \label{eq:CHE_GRS}
\end{align}
and yields all necessary information about properties of local solutions at the singularities of the differential equation: Around both regular singularities local Frobenius-type solutions \cite{Frobenius1873} can be constructed using the entries of the first and second column of the GRS and we can construct Thom\'{e}-type solutions (asymptotic series) at the irregular singularity with the help of the third column.
\subsubsection*{Frobenius solutions}
The two local Frobenius solutions around the regular singularity at $z=0$ are 
\begin{subequations}
\begin{alignat}{2}
y^{I}(z;0) &= & &\sum_{k=0}^\infty a_k \, z^k  \, ,\\
y^{II}(z;0) & = & & \sum_{k=0}^\infty b_k \, z^{k+1-\gamma}
\end{alignat}
\end{subequations}
and those around the other regular singularity at $z=1$ are given by
\begin{subequations}
\begin{alignat}{2}
y^{I}(z;1) &= & &\sum_{k=0}^\infty c_k \, (z-1)^k \, ,\\
y^{II}(z;1) &= & & \sum_{k=0}^\infty d_k \, (z-1)^{k+1-\delta} \, .
\end{alignat}
\end{subequations}
The series expansion coefficients $a_k,b_k,c_k$ and $d_k$ can be obtained by inserting the respective solutions from above into the CHE \eqref{eq:CHE} and solving the resulting three term recurrence relations with chosen initial conditions. Here and in what follows, we use the following notation: The superscript is either $I$ or $II$ to distinguish the two independent local solutions, while the corresponding singularity, at which the solution is constructed, is given in the argument after the semicolon. The local Frobenius solutions $y^{I,II}(z;\cdot)$ are convergent within a circle in the complex plane, centered at the respective regular singularity with a radius that is the distance to the next neighboring singular point. Hence, in each case the region of convergence is bounded by the unit circle centered at the considered singularity.
We emphasize that at each regular singularity $z=z_j \in \left\lbrace 0,1 \right\rbrace$ the solutions have the asymptotic behavior
\begin{align*}
\sim (z-z_j)^0 = \text{const.} \quad \text{or} \quad \sim (z-z_j)^\lambda \quad \text{for } z \to z_j \, ,
\end{align*}
where the (complex) exponent $\lambda$ is either $1-\gamma$ or $1-\delta$, depending on which of the two singularities is considered. Another very useful canonical form of the CHE is obtained by choosing the coefficient functions in \eqref{eq:ODE} to be
\begin{align}
p_1(z) =  a + \dfrac{b+1}{z} + \dfrac{c+1}{z-1} \, , \quad p_0(z) = \dfrac{\mu}{z} + \dfrac{\nu}{z-1} \, , \label{eq:CHE_MapleForm}
\end{align}
and, additionally, defining the two parameters $d$ and $e$ by
\begin{subequations}
\begin{align}
\mu &= \dfrac{1}{2} \left( a-b-c-2e+ab-bc \right) \, ,\\
\nu &= \dfrac{1}{2} \left( a+b+c+2d+2e+ac+bc \right) \, .
\end{align}
\end{subequations}
This form will be called the Maple-form of the CHE in the following, since it is implemented in the computer algebra system Maple  \cite{MapleHeun}.
The GRS for the differential equation in the form \eqref{eq:CHE_MapleForm} then changes to
\begin{align}
\begin{pmatrix}
1 & 1 & 2 &  \\ 
0 & 1 & \infty & ;z \\ 
0 & 0 & \frac{\mu+\nu}{a} & \\ 
-b & -c & b+c+2 - \frac{\mu+\nu}{a} &  \\ 
 &  & 0 &  \\ 
 &  & -a & 
\end{pmatrix} \label{eq:GRS_CHE_MapleForm}
\end{align}
and yields the relations between the parameter sets $\left\lbrace \alpha, \beta, \gamma, \delta, q\right\rbrace$ and $\left\lbrace a, b, c, \mu, \nu \right\rbrace$.
At a first glimpse, this looks like an unnecessary modification but it will prove very useful in the following. We define the standard confluent Heun function $\mathrm{HeunC}(a,b,c,d,e,z)$ in the domain $|z| < 1$ as the first local Frobenius solution at the regular singularity $z = 0$ \cite{MapleHeun}
\begin{align}
\mathrm{HeunC}(a,b,c,d,e,z) := y^{I}(z;0) = \sum_{k=0}^\infty a_k \, z^k
\end{align}
with the additional normalization
\begin{subequations}
\begin{align*}
&~ \mathrm{HeunC}(a,b,c,d,e,z)_{z=0} &&= 1 ·\, , \\[5pt]
&\left. \diff{\mathrm{HeunC}(a,b,c,d,e,z)}{z} \right|_{z=0} &&= \dfrac{b+(c-a)(b+1)+2e}{2(b+1)} \, .
\end{align*}
\end{subequations}
From \eqref{eq:CHE_MapleForm} we obtain the series coefficients $a_k$ as solutions of the recurrence relation
\begin{subequations}
\begin{align}
&a_k = 0 \; \forall k<0 \\
&a_0 = 1 \\
A_k \, &a_k = B_k \, a_{k-1} + C_k \, a_{k-2} \, ,
\end{align}
\end{subequations}
depending on the five parameters $a,b,c,d$ and $e$, where
\begin{subequations}
\label{eq:CHE_recurrence}
\begin{align}
A_k &= 1 + \dfrac{b}{k} \, ,\\
B_k &= 1 - \dfrac{a - b - c +1}{k} - \dfrac{\frac{ab}{2} - \frac{bc}{2} + \frac{b+c-a}{2} - e}{k^2} \, ,\\
C_k &= \dfrac{a}{k} + \dfrac{a}{k^2} \left( \dfrac{d}{a} + \dfrac{b+c}{2} -1 \right) \, .
\end{align}
\end{subequations}
We can read off the behavior for large $k$, since the limit $k \to \infty$ yields
\begin{subequations}
\begin{align*}
\lim_{k \to \infty} A_k = \lim_{k \to \infty} B_k  = 1\, , \lim_{k \to \infty} C_k = 0 \\
\Rightarrow \, a_k -  a_{k-1} \to 0  \, .
\end{align*}
\end{subequations}
Hence, we conclude that the standard solution $\mathrm{HeunC}(a,b,c,d,e,z)$ is convergent and, thus, well defined only for $|z|<1$ as it should be for a local Frobenius solution; the radius of convergence is given by the distance to the next neighboring singularity.
The major advantage is now that once the standard function $\mathrm{HeunC}$ is implemented together with the recurrence relation \eqref{eq:CHE_recurrence}, the second Frobenius solution at $z=0$ and even those two at the other regular singularity $z=1$ can be expressed using this single function \cite{MapleHeun}. The second Frobenius solution at $z=0$ can be represented as
\begin{align}
y^{II}(z;0) = z^{-b} \, \mathrm{HeunC}(a,-b,c,d,e,z)
\end{align}
and the two Frobenius solutions at $z=1$ are now given by
\begin{subequations}
\begin{alignat}{2}
y^{I}(z;1) &= & &\mathrm{HeunC}(-a,+c,b,-d,e+d,1-z) \, ,\\
y^{II}(z;1) &= (z-1)^{-c} \, & &\mathrm{HeunC}(-a,-c,b,-d,e+d,1-z) \, .
\end{alignat}
\end{subequations}
We will need these representations to construct solutions to the radial part of wave equations on Schwarzschild background spacetime in sections \ref{sec:radialEqn1} and \ref{sec:radialEqn2}.
\subsubsection*{Thom\'{e} solutions}
The Thom\'{e} solutions (asymptotic series) at the irregular singularity $z=\infty$ are constructed as \cite{SlavyanovLayBook}
\begin{subequations}
\begin{alignat}{2}
y^{I}(z;\infty) &= & &\sum_{k=0}^{\infty} \rho_k \, z^{-(k+\alpha)} \, , \\
y^{II}(z;\infty) &= \mathrm{e}^{\beta z} & & \sum_{k=0}^{\infty} \sigma_k \, z^{-(k+\gamma+\delta-\alpha)}
\end{alignat}
\end{subequations}
and the coefficients $\rho_k$ and $\sigma_k$ follow, again, from recursion relations that can be derived by inserting $y^{I,II}(z;\infty)$ into the differential equation.
\subsection{\label{subsec:waveeq} Wave equations in Schwarzschild spacetime}
For the entire article we choose natural units such that the speed of light, Newton's Gravitational constant, Planck's constant and Boltzmann's constant are set to unity, i.e., $c=G=\hbar=k_B=1$. Furthermore, all distances are measured in multiples of the Schwarzschild radius $r_s = 2M$. Using this convention, the horizon is located at $r=1$ and the light sphere at $r=3/2$, where $r$ is the usual Schwarzschild radial coordinate (i.e., the areal radius). The Schwarzschild metric in these units is 
then
\begin{align*}
 g_{\mu\nu} dx^\mu dx^\nu = - \dfrac{(r-1)}{r} dt^2 + \dfrac{r}{(r-1)} dr^2 + r^2 d\Omega^2 \, ,
\end{align*}
with $d\Omega^2$ being the standard metric on the unit two-sphere.
\subsubsection*{Tortoise coordinates}
Since it will become very useful later, we start with an analysis of the wave equation 
\begin{align}
\left[ \dfrac{\partial^2}{\partial r_*^2} - \diffp[2]{}{t} - V(r) \right] \Psi(t,r_*) = 0 \label{eq:waveEqn_general}
\end{align}
that involves the Schwarzschild time coordinate $t$ and the tortoise coordinate 
\begin{subequations}
\label{eq:tortoiseCoord}
\begin{align}
r_* &= r+\log(r-1) \, , \\
\mathrm{d}r_* &= \diff{r_*}{r} ~ \mathrm{d}r = \dfrac{r}{(r-1)}  \mathrm{d}r \, .
\end{align}
\end{subequations}
Here $V(r)$ is some given spherically symmetric potential. For $V(r)=0$ we obtain the free wave equation, which is solved by
\begin{align}
\Psi(t,r_*) = A(\omega) \, \mathrm{e}^{-i \omega (t+r_*)} + B(\omega) \, \mathrm{e}^{-i \omega (t-r_*)} \, ,
\end{align}
i.e., by a superposition of in- and outgoing waves with arbitrary complex amplitudes $A(\omega)$ and $B(\omega)$.
In the following we transform the wave equation \eqref{eq:waveEqn_general} into the corresponding wave equations in usual Schwarzschild, Eddington-Finkelstein (EF), Painlev\'{e}-Gullstrand (PG) and Kruskal-Szekeres (KS) coordinates.
\subsubsection*{Schwarzschild coordinates}
To obtain the wave equation in usual Schwarzschild coordinates we perform the simple transformation $r_* \mapsto r$ that does not affect any of the other coordinates. Using \eqref{eq:tortoiseCoord} and the chain rule we rewrite 
\begin{subequations}
\begin{align}
\diffp{}{{r_*}} &= \diff{r}{{r_*}} \diffp{}{r} = \dfrac{(r-1)}{r} \diffp{}{r} \\
\Rightarrow \diffp[2]{}{{r_*}} &= \dfrac{(r-1)}{r}  \left[ \dfrac{(r-1)}{r} \diffp[2]{}{r} + \dfrac{1}{r^2} \diffp{}{r} \right] 
\end{align}
\end{subequations}
and the wave equation \eqref{eq:waveEqn_general} becomes now
\begin{multline}
\left[ \left( \dfrac{r-1}{r} \right)^2 \diffp[2]{}{r} + \dfrac{(r-1)}{r^3} \diffp{}{r} \right. \\
- \left. \diffp[2]{}{t} - V(r) \right] \Psi(t,r) = 0 \, . \label{eq:waveEqn_Schwarzschild}
\end{multline}
This equation contains a ``new'' first-derivative term and we notice that the values $r=0$ and $r=1$, i.e., the origin and the horizon in our coordinates, describe singularities of this differential equation. Starting from \eqref{eq:waveEqn_Schwarzschild} we use a further coordinate transformation $(t,r) \mapsto (u,v)$, where
\begin{align}
v = f(r,t) \, , \quad u = g(r,t) \, , \label{eq:coordTrafo}
\end{align}
to consider additional coordinate systems. The angles $\vartheta$ and $\varphi$ remain unchanged to keep the coordinates adapted to spherical symmetry. In the wave equation \eqref{eq:waveEqn_Schwarzschild} partial derivatives with respect to $r$ are meant that $t$ is
kept fixed, and vice versa. In the new coordinates, we have to work with partial derivatives with respect to $u$ where $v$ is kept fixed, and vice versa. Thus, we need the following transformation formulas for partial derivatives:
\begin{subequations}
\begin{align}
\left. \diffp{}{t} \right|_{r=\text{const.}} &= \left. \diffp{v}{t} \diffp{}{v} \right|_{u=\text{const.}} + \left. \diffp{u}{t} \diffp{}{u} \right|_{v=\text{const.}} \notag
\\[6pt]
&=: \dot{f}(r,t) \diffp{}{v} + \dot{g}(r,t) \diffp{}{u} \, ,\\[12pt]
\left. \diffp{}{r} \right|_{t=\text{const.}} &= \left. \diffp{v}{r} \diffp{}{v} \right|_{u=\text{const.}} + \left. \diffp{u}{r} \diffp{}{u} \right|_{v=\text{const.}} \notag
\\[6pt]
&=: f'(r,t) \diffp{}{v} + g'(r,t) \diffp{}{u} \,  .
\end{align}
\end{subequations}
If, for the sake of readability, we omit the arguments of the functions $f$ and $g$, the wave equation reads
\begin{align}
& \left[ \left(\dfrac{r-1}{r}\right)^2 {f'}^2 - \dot{f}^2 \right] \diffp[2]{\Psi(u,v)}{v} \notag \\
+& \left[ \left(\dfrac{r-1}{r}\right)^2 {g'}^2 - \dot{g}^2 \right] \diffp[2]{\Psi(u,v)}{u} \notag \\
+& \left[ 2\left( \dfrac{r-1}{r}\right)^2 f' g' - 2 \dot{f} \dot{g} \right] \diffp{\Psi(u,v)}{u v} \notag \\
+& \left[ \left( \dfrac{r-1}{r} \right)^2 f'' + \dfrac{(r-1)}{r^3} f' - \ddot{f} \right] \diffp{\Psi(u,v)}{v} \notag \\
+& \left[ \left( \dfrac{r-1}{r} \right)^2 g'' + \dfrac{(r-1)}{r^3} g' - \ddot{g} \right] \diffp{\Psi(u,v)}{u} \notag \\
-& V(r) \Psi(u,v)  = 0 \, ,\label{eq:waveEqn_coordTrafo}
\end{align}
where $r$ has to be considered as a function of the new coordinates, i.e., $r = r(u,v)$. Now, we can introduce Eddington-Finkelstein, Painlev\'{e}-Gullstrand and Kruskal-Szekeres coordinates by specifying the two functions $f(r,t)$ and $g(r,t)$, respectively. For an overview of such regular coordinate systems for the Schwarzschild spacetime the reader is referred to, e.g., the work in \cite{MartelPoisson2001} and references therein.
\subsubsection*{Eddington-Finkelstein coordinates}
For (ingoing) Eddington-Finkelstein coordinates the proper coordinate transformation is given by
\begin{subequations}
\label{eq:coordTrafo_EF}
\begin{align}
v &= t + r + \log(r-1) \, , \\
u & \equiv r \, .
\end{align}
\end{subequations}
The new time coordinate $v$ is constant on ingoing radial null geodesics. With the transformation above the wave equation \eqref{eq:waveEqn_coordTrafo} reduces to the simpler form
\begin{multline}
\left[ \left(\dfrac{r-1}{r}\right)^2 \diffp[2]{}{r} + 2  \dfrac{(r-1)}{r} \diffp{}{r v} \right. \\
\left. + \dfrac{(r-1)}{r^3} \diffp{}{r} -V(r) \right] \Psi(v,r) = 0 \, . \label{eq:waveEqn_EF}
\end{multline}
\subsubsection*{Painlev\'{e}-Gullstrand coordinates}
To consider Painlev\'{e}-Gullstrand coordinates we have to use instead the slightly more complicated coordinate transformation
\begin{subequations} 
\label{eq:coordTrafo_PG}
\begin{eqnarray}
v &=& t + 2\sqrt{r} + \log \left| \dfrac{\sqrt{r}-1}{\sqrt{r}+1} \right| \, ,\\
u &\equiv& r \, .
\end{eqnarray} 
\end{subequations}
Calculating all necessary first and second derivatives we obtain
\begin{multline}
\left[ \left( \dfrac{r-1}{r} \right)^2 \diffp[2]{}{r} - \dfrac{(r-1)}{r} \diffp[2]{}{v}
+ 2 \dfrac{(r-1)}{r^{3/2}} \diffp{}{r v} \right. \\ \left. - \dfrac{(r-1)}{2 r^{5/2}}  \diffp{}{v}
 + \dfrac{(r-1)}{r^3} \diffp{}{r} - V(r) \right] \Psi(v,r) = 0 \label{eq:waveEqn_PG}
\end{multline}
as the corresponding wave equation in PG coordinates. Its qualitative features are similar to those of the wave equation in EF coordinates. However, as the latter looks simpler we will concentrate
in the following on the EF coordinates.
 \subsubsection*{Kruskal-Szekeres coordinates}
The transformation to Kruskal-Szekeres coordinates (in the outer region $r>1$) is achieved by 
\begin{subequations}
\label{eq:coordtrafoKS}
\begin{align}
v &= \sqrt{r-1} ~ \mathrm{e}^{r/2} \sinh(t/2)  \, ,\\
u &= \sqrt{r-1} ~ \mathrm{e}^{r/2} \cosh(t/2) 
\end{align}
\end{subequations}
and, thus, equation \eqref{eq:waveEqn_coordTrafo} simplifies to
\begin{align}
\left[ (r-1) \left(\diffp[2]{}{u} - \diffp[2]{}{v} \right)  -4 \mathrm{e}^{-r} V(r) \right] \Psi(u,v) = 0 \, . \label{eq:waveEqn_KS}
\end{align}
Here, $r$ has to be considered as a function of $u$ and $v$, implicitly given by the relation
\begin{align}
(r-1) \, \mathrm{e}^{r} = u^2 - v^2 \, ,
\end{align}
and we recognize the wave operator in double null coordinates $\partial^2/\partial u^2 - \partial^2/\partial v^2$. In the interior region $(r<1)$ $u$ and $v$ have to be interchanged and this corresponds to a change of signs in the wave operator. While in the four other coordinate systems the wave equation can be reduced with a separation ansatz to a stationary equation by splitting off a time factor, this is not true in KS coordinates. For this reason, we will not consider the KS coordinates in  the rest of the paper.
\subsection{Linear wave equations for scalar, electromagnetic and gravitational perturbations}
For each (massless) perturbation with any of the spin values $s=0,1,2$ a wave equation of the form \eqref{eq:waveEqn_general} can be derived. These wave equations are valid in the sense that the back reaction on the spacetime geometry is neglected and we do consider the fields as small perturbations on a fixed background spacetime. In the following, we sketch how to derive the respective wave equation on Schwarzschild background for a chosen spin value and combine the results in the end to obtain the general Regge-Wheeler equation that contains the spin as a parameter.
\subsubsection*{Scalar perturbations}
For (massless) scalar perturbations the wave equation is a result of the Klein-Gordon-equation 
\begin{align}
\Box \Phi \equiv \nabla^\mu \nabla_\mu \Phi = \sqrt{-g} \dfrac{\partial}{\partial x^\mu}\left[ \sqrt{-g} ~ g^{\mu\nu} \dfrac{\partial}{\partial x^\nu} \right] \Phi = 0 \, , \label{eq:KleinGordon}
\end{align}
where $g$ is the determinant of the Schwarzschild metric. (Note that we use the signature convention (-,+,+,+).) If we expand the field into scalar spherical harmonics $Y_{l m}(\vartheta,\varphi)$ according to \cite{FrolovNovikovBook}
\begin{align}
\Phi = \Phi(t,r_*,\vartheta,\varphi) = \sum^{\infty}_{l=0} \sum^{l}_{m= -l} \dfrac{\Psi_l(t,r_*)}{r} Y_{lm}(\vartheta,\varphi)
\end{align}
and insert this ansatz into the Klein-Gordon equation, we notice that the remaining function $\Psi_l(t,r_*)$ has to fulfill the wave equation \eqref{eq:waveEqn_general} with a potential given by
\begin{align}
V_{l,s=0}(r) = \dfrac{(r-1)}{r^3} \left( l(l+1)+ \dfrac{1}{r} \right) \, . \label{eq:potential_Scalar}
\end{align}
\subsubsection*{Electromagnetic perturbations}
To describe the propagation of electromagnetic waves we consider the source-free Maxwell equations
\begin{subequations}
\begin{align}
\nabla_\mu F^{\mu\nu} &= 0 \, , \label{eq:Maxwell1} \\ 
\epsilon^{\mu\nu\rho\sigma} \nabla_\nu F_{\rho\sigma} &= 0 \, . \label{eq:Maxwell2}
\end{align}
\end{subequations}
Owing to the conformal invariance of \eqref{eq:Maxwell1} and \eqref{eq:Maxwell2} we can use the metric
\begin{align}
\tilde{g}_{\mu\nu} dx^\mu dx^\nu =-dt^2 + dr_*^2 + \dfrac{r^3 \left(d\vartheta^2 + \sin^2 \vartheta  d\varphi^2 \right)}{(r-1)} \, , \label{eq:conformalmetric}
\end{align}
which is conformally equivalent to the Schwarzschild metric. Following Stephani \cite{Stephani1974}, the source-free Maxwell equations can be reduced to the Debye-equation, which is a single differential equation for a scalar function called Debye potential. For the background spacetime described by \eqref{eq:conformalmetric} this Debye equation is
\begin{multline}
\left[ \frac{(r-1)}{r^3} \left( \cot\vartheta \diffp{}{\vartheta} + \diffp[2]{}{\vartheta} + \dfrac{1}{\sin^2 \vartheta} \diffp[2]{}{\varphi} \right) \right. \\
\left. + \dfrac{\partial^2}{\partial r_*^2} - \diffp[2]{}{t} \right] \Pi = 0 \, , \label{eq:DebyeEqn}
\end{multline}
which is given in \onlinecite{HerltStephani1975} as well. To obtain the general solution of Maxwell's equations we have to construct the four-potential of the electromagnetic field \cite{Stephani1974}
\begin{align}
A_\mu = (u^\nu v_\mu - v^\nu u_\mu) \partial_\nu \Pi + \epsilon_\mu^{~\nu\lambda\sigma} v_\lambda u_\sigma \partial_\nu \Phi \, ,
\end{align}
where $\Pi$ and $\Phi$ are two independent solutions of the Debye-equation and $(v^\mu) = (0,1,0,0) \, , \, (u^\mu) = (1,0,0,0)$. Thereupon, we can obtain the components of the electromagnetic field as usual by
\begin{align}
F_{\mu\nu} = \partial_\mu A_\nu - \partial_\nu A_\mu \, .
\end{align}
Since the Debye potential $\Pi$ in \eqref{eq:DebyeEqn} is a scalar function, we can expand it into scalar spherical harmonics according to
\begin{align}
\Pi(t,r_*,\vartheta,\varphi) = \sum^{\infty}_{l=1} \sum^{l}_{m= -l} \Psi_l(t,r_*) Y_{lm}(\vartheta,\varphi) \, .
\end{align}
Inserting this ansatz into the Debye equation, we notice that the function $\Psi_l(t,r_*)$ has  to fulfill the wave equation \eqref{eq:waveEqn_general} with a potential given by
\begin{align}
V_{l,s=1}(r) = \dfrac{(r-1)}{r^3} l(l+1) \, . \label{eq:potential_EM}
\end{align}
\subsubsection*{Gravitational perturbations}
In a linear perturbation approach, as considered by Regge and Wheeler \cite{ReggeWheeler1957}, we write the total metric $g_{\mu \nu}$ as the sum 
\begin{align*}
g _{\mu \nu} = \overline{g} {}_{\mu \nu} + h _{\mu \nu}
\end{align*}
of a background metric $\overline{g} _{\mu \nu}$ and a small perturbation $h _{\mu \nu}$. We neglect all higher order terms in $h_{\mu \nu}$ and its derivatives. A rather lengthy calculation, involving the expansion of the perturbation in scalar, vectorial and tensorial spherical harmonics and restricting to odd parity perturbations yields the wave equation \eqref{eq:waveEqn_general} with the potential
\begin{align}
V_{l,s=2}(r) = \dfrac{(r-1)}{r^3} \left( l(l+1) - \dfrac{3}{r} \right) \label{eq:potential_Grav}
\end{align}
as a result of the Einstein vacuum field equation. The (odd-parity) components of the perturbation $h_{\mu \nu}$ can be calculated in terms of the solutions to this wave equation. A similar treatment for the even parity perturbations leads to the Zerilli equation \cite{ChandrasekharBook} which, however will not be considered in the present paper. 
\subsubsection*{The general Regge-Wheeler equation}
Combining the previous results for different spins we obtain the general time-dependent Regge-Wheeler equation
\begin{align}
\left[ \diffp[2]{}{r_*} - \diffp[2]{}{t} - V_{s l}(r) \right] \Psi_{s l}(t,r_*) = 0 \, , \label{eq:RWE_time}
\end{align}
where the spin-dependent potential is now given by
\begin{align}
V_{s l}(r) = \dfrac{(r-1)}{r^3} \left( l(l+1) + \dfrac{1-s^2}{r} \right) \, . \label{eq:RWE_potential}
\end{align}
The differential equation \eqref{eq:RWE_time} is a wave equation of the form \eqref{eq:waveEqn_general} and can, thus, be transformed to the corresponding wave equations in Schwarzschild, Eddington-Finkelstein, Painlev\'{e}-Gullstrand and Kruskal-Szekeres coordinates  using the results of section \eqref{subsec:waveeq}.
\section{\label{sec:radialEqn1} Radial equation in tortoise and Schwarzschild coordinates}
\subsection{Stationary Regge-Wheeler equation}
\begin{figure}
\begin{center}
\includegraphics[width=0.4\textwidth]{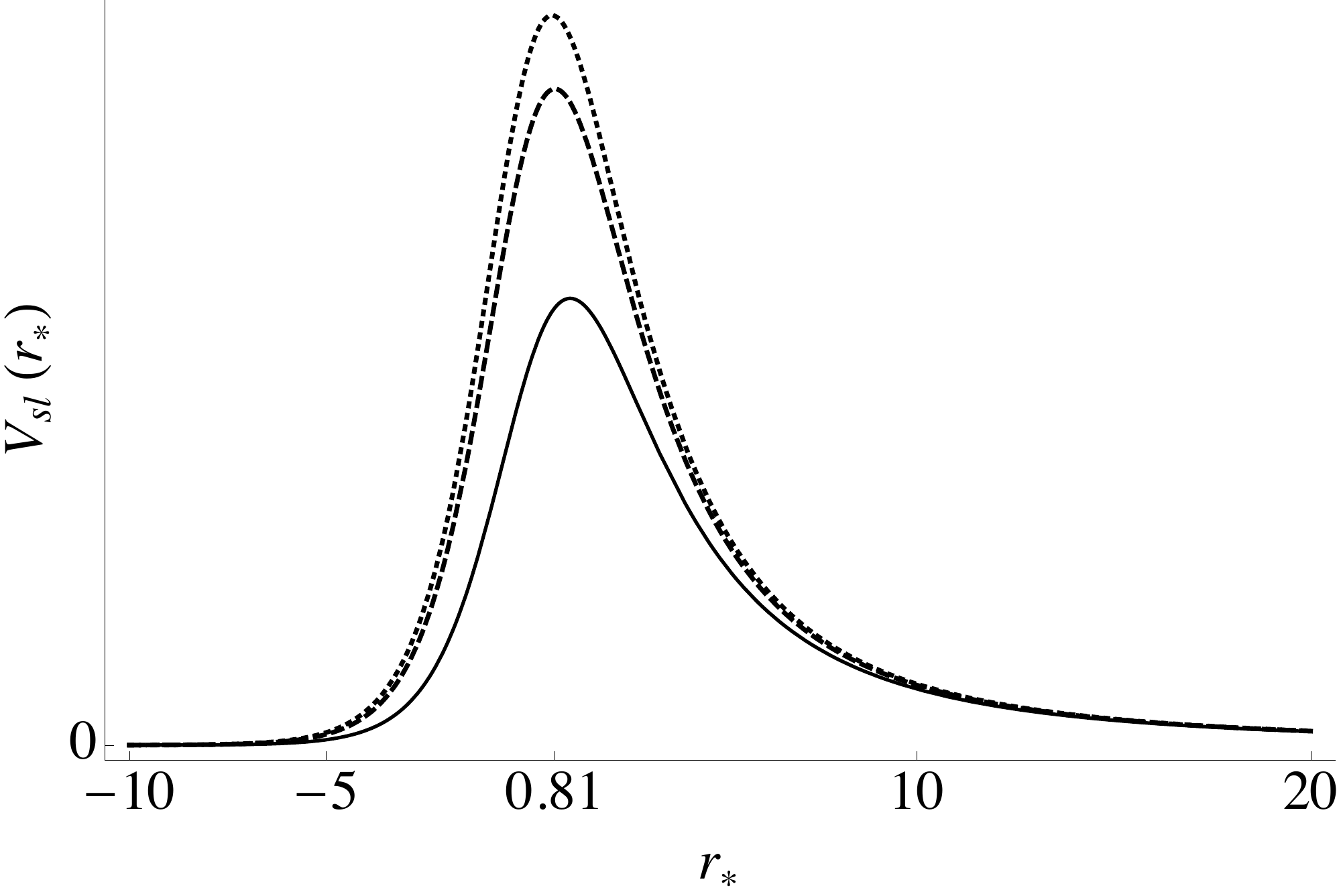}
\end{center}
\caption{\label{fig:RWEpotential} The Regge-Wheeler potential as a function of the tortoise coordinate $r_*$ for $l=2$ and spin $s=0$ (dotted), $s=1$ (dashed) and $s=2$ (solid).}
\end{figure}

\begin{figure}
\begin{center}
\includegraphics[width=0.4\textwidth]{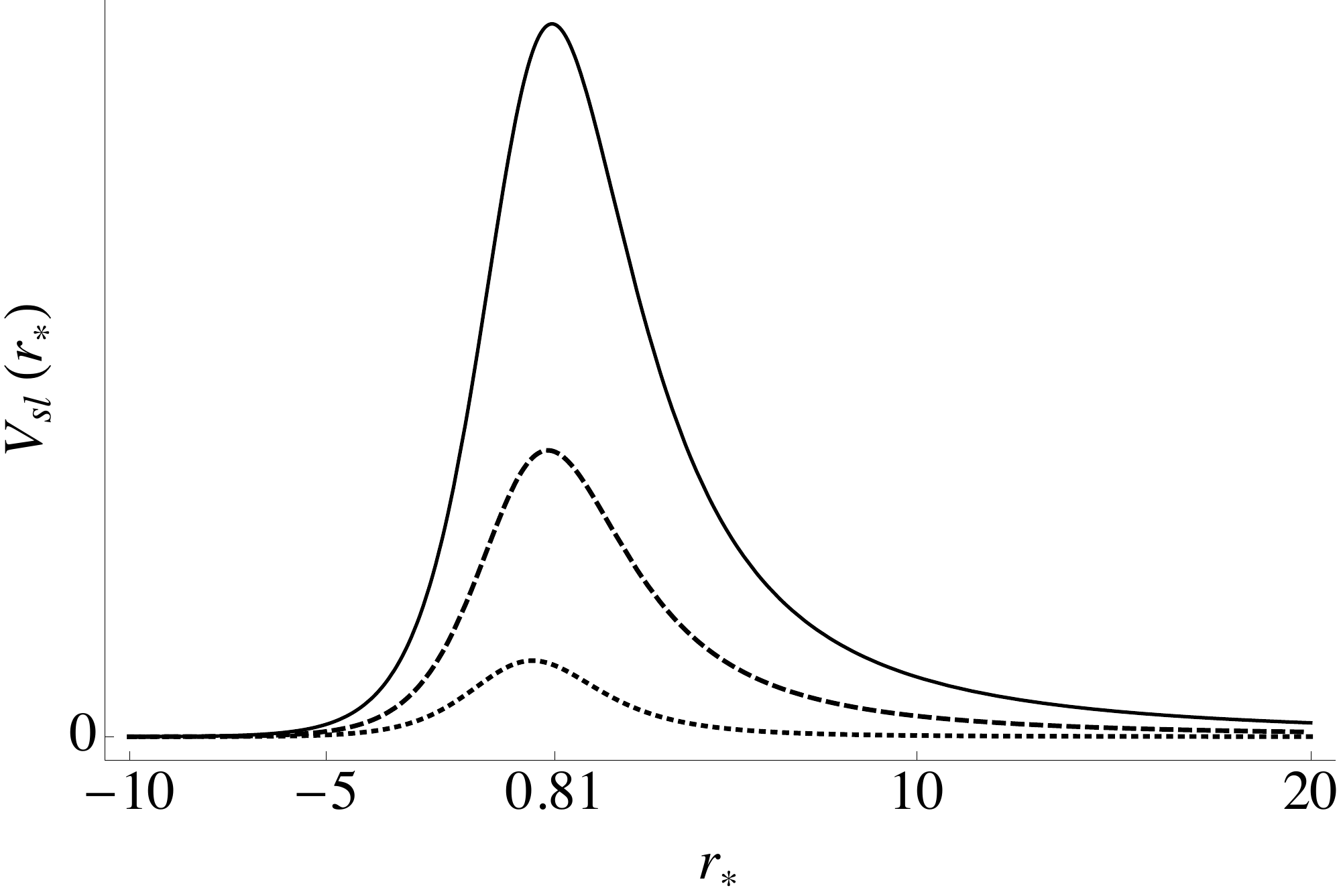}
\end{center}
\caption{\label{fig:RWEpotential2} The Regge-Wheeler potential as a function of the tortoise coordinate $r_*$ for $s=0$ and $l=0$ (dotted), $l=1$ (dashed) and $l=2$ (solid).}
\end{figure}

A lot of the phenomenology can be understood by investigating the wave equation in the simplest form, i.e., written in the tortoise coordinate. Hence, we shall briefly recap the properties of this differential equation. Assuming a harmonic time dependence
\begin{align}
\Psi_{s l}(t,r_*) = \mathrm{e}^{-i \omega t} R_{\omega s l}(r_*)
\end{align}
in \eqref{eq:RWE_time} leads to the stationary form of the Regge-Wheeler equation (RWE) that is
\begin{align}
\left[ \diff[2]{}{r_*} + \omega^2 - V_{s l}(r) \right] R_{\omega s l}(r_*) = 0 \, . \label{eq:RWE}
\end{align}
This radial equation is an ordinary, second order, linear and homogeneous differential equation in the form of a Schr\"odinger equation with ``energy'' $\omega^2$. 
The shape of the involved potential is shown in Fig. \eqref{fig:RWEpotential} for all three spin values and a partial wave index $l=2$ as an example.
The maxima occur near the lightsphere radius $r=3/2$ ($r_* \approx 0.81$). While for $s=1$ the maximum is always located exactly at that radius, for $s=0$ it is found slightly below and for 
$s=2$ slightly above, but in any case it coincides with the lightsphere radius in the limit $l \to \infty$. The value of the potential increases with the partial wave number (angular momentum index). Figure (\ref{fig:RWEpotential2}) shows this $l$-dependence of the maximum for $s=0$.
The potential decreases exponentially near the horizon ($r_* \to -\infty$) and behaves like $r^{-2}$ for $r_* \to +\infty$. Hence, the asymptotic solutions in these limits are 
\begin{align}
R_{\omega s l}(r_*) =
\begin{cases}
&A_{\omega s l} \, \mathrm{e}^{-i \omega r_*}  + B_{\omega s l} \, \mathrm{e}^{i \omega r_*}, \quad r_* \to -\infty \notag \\
&C_{\omega s l} \, \mathrm{e}^{-i \omega r_*}  + D_{\omega s l} \, \mathrm{e}^{i \omega r_*}, \quad r_* \to +\infty
\end{cases} \, .
\end{align}
This is, together with the time-dependent factor $\mathrm{e}^{-i \omega t}$, a superposition of in- and outgoing waves, where terms $\sim \mathrm{e}^{-i\omega r_*}$ describe waves that propagate to smaller radii and terms $\sim \mathrm{e}^{i\omega r_*}$ describe waves propagating to larger radii with increasing time.
To consider a black hole without quantum effects we have to assume causal boundary conditions to ensure that no waves emerge from the horizon. Thus, the asymptotic solution that satisfies this causal boundary condition is the IN-mode  \cite{FrolovNovikovBook}
\begin{align}
\label{eq:INmode}
R_{\omega s l}^{IN}(r_*) = 
\begin{cases}
A^{trans}_{\omega l} \, \mathrm{e}^{-i\omega r_*}, \quad &r_* \to -\infty \\
A^{in}_{\omega l} \, \mathrm{e}^{-i \omega r_*} + A^{out}_{\omega l} \, \mathrm{e}^{i \omega r_*}, \quad &r_* \to +\infty
\end{cases}  \, .
\end{align}
Usually, the amplitude at the horizon is normalized such that $A^{trans} =1$ and we can define the reflection and transmission coefficients \cite{FrolovNovikovBook}
\begin{align}
\mathbb{T}_{\omega l} = |T|^2_{\omega l} = \left| \dfrac{1}{A^{in}_{\omega l}} \right|^2, \quad \mathbb{R}_{\omega l} = |R|^2_{\omega l} = \left| \dfrac{A^{out}_{\omega l}}{A^{in}_{\omega l}} \right|^2 \, .
\end{align}
The complex conjugate of \eqref{eq:INmode} gives the OUT-mode, $R_{\omega s l}^{OUT}(r_*)$.  
Considering the Wronskian $W(R^{IN},R^{OUT})$, we can derive the relation
\begin{align}
\mathbb{T}_{\omega l}  + \mathbb{R}_{\omega l} = 1
\end{align}
that corresponds to flux conservation.
To consider scattering at a black hole we have to find a solution of the RWE \eqref{eq:RWE} that satisfies the causal boundary condition, i.e., that has the asymptotic behavior \eqref{eq:INmode}. From this solution we could read off the amplitudes $A^{in}$ and $A^{out}$ to obtain the reflection and transmission coefficients or the phase shifts $\delta_l$
\begin{align}
\mathrm{e}^{2i \delta_l} \sim \dfrac{A^{out}_{\omega l}}{A^{in}_{\omega l}} \, ,
\end{align}
which give rise to the (total) scattering and absorption cross sections of the black hole. 

To obtain quasi-normal modes (QNM) and calculate their frequencies one usually restricts to purely ingoing boundary conditions: No waves emerge from the horizon, but no waves are coming in from spatial infinity as well. Hence, an initial perturbation radiates into the horizon and into spatial infinity and should decay in the cause of time in order to have a black hole that is stable against the considered perturbation. This leads, of course, to imaginary frequencies, where the real part describes an ordinary oscillation and the imaginary part (with the proper sign) assures the temporal decay.
The proper asymptotic solution is then given by
\begin{align}
R_{\omega s l}^{QNM}(r_*) = 
\begin{cases}
A^{QNM}_{\omega l} \, \mathrm{e}^{-i\omega r_*}, \quad &r_* \to -\infty \\
B^{QNM}_{\omega l} \, \mathrm{e}^{i \omega r_*}, \quad &r_* \to +\infty
\end{cases} \, .
\end{align}
Unfortunately, no exact solution of the RWE \eqref{eq:RWE} is known due to the involved potential \eqref{eq:RWE_potential}. Thus, to solve the scattering or quasi-normal mode problem the only possibilities seem to be either approximation methods (e.g., WKB \cite{HerltStephani1975}, Born \cite{Batic2011}, ...) or numerical solutions. But yet another method is shown in the next sections where we derive exact analytic solutions, given in the form of convergent series, after changing the coordinate system and thus the underlying wave equation.
\subsection{Radial equation in Schwarzschild coordinate}
As a result of section \ref{subsec:waveeq} we know that the analog to the time-dependent RWE (coordinates $t$ and $r_*$) in Schwarzschild coordinates ($t$ and $r$) is obtained by using equation \eqref{eq:waveEqn_Schwarzschild} together with the Regge-Wheeler potential \eqref{eq:RWE_potential}. We can reduce this wave equation, again, to a radial equation with the ansatz
\begin{align*}
\Psi_{s l}(t,r) = \mathrm{e}^{-i \omega t} R_{\omega s l}(r)
\end{align*}
as was done before to derive the stationary RWE. This leads to the radial equation
\begin{multline}
\left[ r(r-1)^2 \diff[2]{}{r} + (r-1) \diff{}{r} \right.
+ r^3 \omega^2 \\ 
\left. - (r-1) \left( l(l+1) + \dfrac{1-s^2}{r} \right) \right] R_{\omega s l}(r) = 0 \, , \label{eq:radialeqnSchwarzschild}
\end{multline}
which is discussed in the literature for example by Leaver \cite{Leaver1985, Leaver1986} and by Fiziev \cite{Fiziev2006, Fiziev2007}. Leaver uses the ansatz
\begin{align}
R_{\omega s l}(r) = r^{s+1} (r-1)^{-i\omega} y_{\omega s l}(r)
\end{align}
to transform equation (\ref{eq:radialeqnSchwarzschild}) into a generalized spheroidal wave equation for $ y_{\omega s l}(r)$, which is nothing but a special form of the confluent Heun equation (CHE), even though Leaver never stated this nor did he use the term ``confluent Heun equation'' in his work. Fiziev on the other hand makes the ansatz
\begin{align}
R_{\omega s l}(r) = r^{s+1} (r-1)^{i\omega} \mathrm{e}^{i \omega r} H_{\omega s l}(r) \label{eq:AnsatzFieziev}
\end{align}
and gives the confluent Heun equation in the Maple-form for the remaining function $H_{\omega s l}(r)$ and its solutions in terms of the standard confluent Heun function. It should be stated that in both cases the ansatz is not well defined at the horizon ($r=1$) due to the term $(r-1)^{\pm i\omega}$. In the following, we will briefly recall these known solutions since we can use them for a comparison of our results later. With the ansatz of Fiziev the function $H_{\omega s l}(r)$ has to satisfy a confluent Heun equation in the Maple-form \eqref{eq:CHE_MapleForm} with the five parameters
\begin{align}
a &= 2i\omega \, , \quad b = 2s \, , \quad c = 2i\omega \, , \\
d &= 2\omega^2 \, , \quad e = s^2 - l(l+1) \, .
\end{align}
The two regular singularities are located at the origin $r=0$ and at the horizon $r=1$, while the irregular singularity is found at spatial infinity $r=\infty$. We can give the GRS of this equation using the general result (\ref{eq:GRS_CHE_MapleForm}) to complement the work in \cite{Fiziev2006} and \cite{Fiziev2007}. This GRS then reads
\begin{align}
\begin{pmatrix}
1 & 1 & 2 &  \\ 
0 & 1 & \infty & ;r \\ 
0 & 0 & s+1 & \\ 
-2s & -2i\omega & s+1 + 2i\omega &  \\ 
 &  & 0 &  \\ 
 &  &-2i\omega&
\end{pmatrix} \, . \label{eq:GRS_Schwarzschild}
\end{align}
Thus, the two local Frobenius solutions at the horizon can easily be constructed manually with the help of the above GRS or given in terms of the standard confluent Heun function as follows:
\begin{subequations}
\begin{alignat}{2}
H^{I}_{\omega s l}(r;1) &= (r-1)^{-c}\, & &\mathrm{HeunC}(-a,-c,b,-d,e+d,1-r) \notag \\
H^{II}_{\omega s l}(r;1) &= & &\mathrm{HeunC}(-a,+c,b,-d,e+d,1-r) . \notag
\end{alignat}
\end{subequations}
Now, we use again (\ref{eq:AnsatzFieziev}) and obtain the full solutions of the Schwarzschild radial perturbation equation (\ref{eq:radialeqnSchwarzschild}) around the regular singularity at the horizon. These solutions are
\begin{widetext}
\begin{subequations}
\begin{alignat}{2}
R^{I}_{\omega s l}(r;1) &= r^{s+1} \mathrm{e}^{i \omega (r-\log(r-1))} & &\mathrm{HeunC}(-2i\omega,-2i\omega,2s,-2\omega^2,2\omega^2 + s^2 - l(l+1),1-r) \, , \\
R^{II}_{\omega s l}(r;1) &= r^{s+1} \mathrm{e}^{i \omega (r+\log(r-1))}  & &\mathrm{HeunC}(-2i\omega,+2i\omega,2s,-2\omega^2,2\omega^2 + s^2 - l(l+1),1-r) \, .
\end{alignat}
\end{subequations}
\end{widetext}
It is important to remark that both solutions are not well defined exactly at the horizon. While the modulus approaches one, the phase is not defined: In the complex plane infinitely many turns on the unit circle are performed when approaching the singularity at $r=1$. Thus, in these coordinates it is not possible to construct wavelike solutions that propagate through the horizon. Furthermore, the representation of the solutions in terms of the $\mathrm{HeunC}$-function is only useful in the domain $|r-1|<1$.

The asymptotic Thom\'{e} solutions at the irregular singularity $r=\infty$ can be constructed, as shown in Sec. \ref{sec:prep}, using the characteristic exponents of the second kind in the third column of the GRS \eqref{eq:GRS_Schwarzschild}
\begin{subequations}
\begin{alignat}{2}
H^{I}_{\omega s l}(r;\infty) &= r^{-s+1} & &\sum^\infty_{k=0} \rho_k \, r^{-k} \, , \notag \\
H^{II}_{\omega s l}(r;\infty) &= r^{-s+1} r^{-2i\omega} \mathrm{e}^{-2i\omega r} & &\sum^\infty_{k=0} \sigma_k \, r^{-k}  \, . \notag
\end{alignat}
\end{subequations}
Thereupon, the full Thom\'{e} solutions of the Schwarzschild radial equation at $z=\infty$ are
\begin{subequations}
\begin{alignat}{2}
R^{I}_{\omega s l}(r;\infty) &= \underbrace{\mathrm{e}^{i\omega (r + \log(r-1))}}_{\approx \, \mathrm{e}^{i\omega r}} & &\sum^\infty_{k=0} \rho_k \, r^{-k} \, ,\\
R^{II}_{\omega s l}(r;\infty) &= \underbrace{\mathrm{e}^{-i\omega (r + \log(r-1) -2\log(r))}}_{\approx \, \mathrm{e}^{-i\omega r}} & &\sum^\infty_{k=0} \sigma_k \, r^{-k}
\end{alignat}
\end{subequations}
which coincide with those given in \cite{Fiziev2006}. These series are convergent in the limit $r \to \infty$. For the expansion coefficients $\rho_k$ and $\sigma_k$ recurrence relations can be derived and the normalizations $\rho_0$ and $\sigma_0$, i.e., the amplitudes at spatial infinity remain free to choose.

\begin{figure}
\begin{center}
\includegraphics[width=0.35\textwidth]{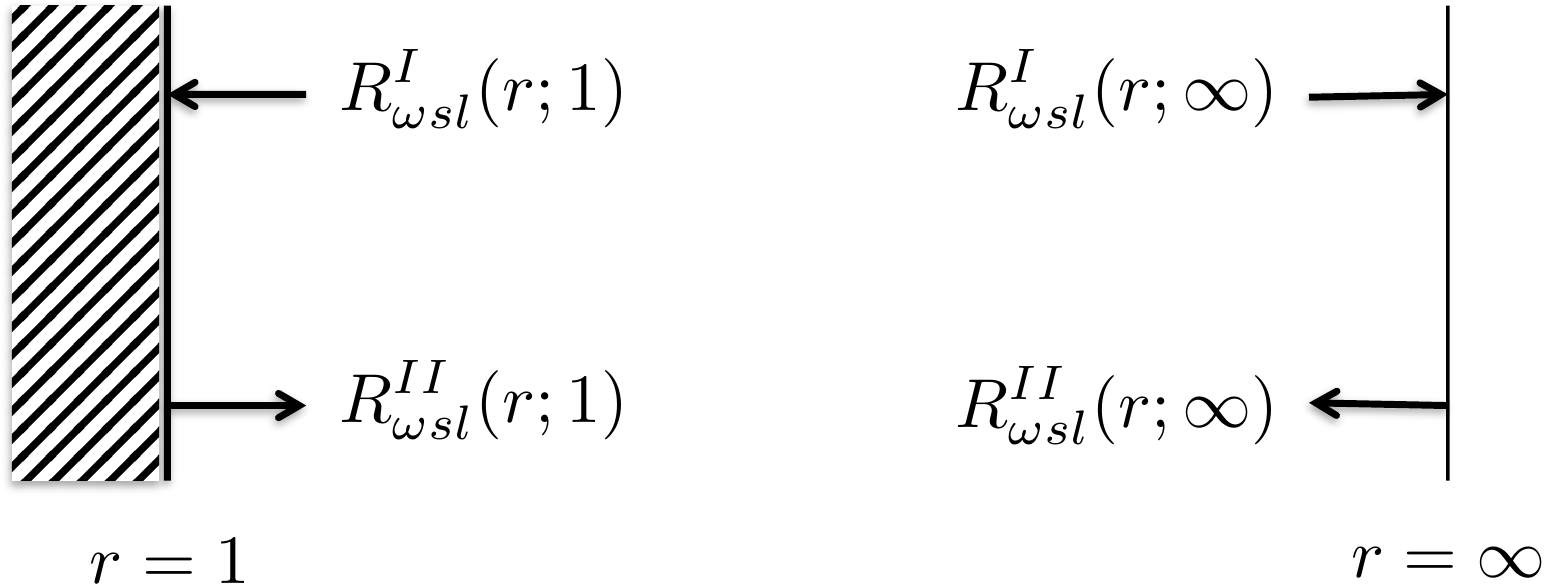}
\end{center}
\caption{\label{fig:InOut_Schwarzschild} In- and outgoing properties of the local solutions to the Schwarzschild and Eddington-Finkelstein radial equations.}
\end{figure}

Combined with the time-dependent factor $\mathrm{e}^{-i\omega t}$ the solutions $R^{I}_{\omega s l}(r;1)$ and $R^{I}_{\omega s l}(r;\infty)$ describe waves that propagate into the respective singularity, i.e., waves going into the horizon and waves going into spatial infinity.
The solutions $R^{II}_{\omega s l}(r;1)$ and $R^{II}_{\omega s l}(r;\infty)$ describe waves that emerge from the respective singular point. Figure (\ref{fig:InOut_Schwarzschild}) sketches these in- and outgoing properties as a mnemonic.
\section{\label{sec:radialEqn2} Radial equation in Eddington-Finkelstein coordinates}
To obtain the radial perturbation equation in Eddington-Finkelstein coordinates we use the wave equation \eqref{eq:waveEqn_EF} together with the Regge-Wheeler potential \eqref{eq:RWE_potential}. Thereupon, we perform the ansatz
\begin{align}
\Psi_{s l}(v, r) = \mathrm{e}^{-i \omega v} R_{\omega s l}(r)
\end{align}
to separate the Eddington-Finkelstein time coordinate $v$. Thus, the remaining function $R_{\omega s l}(r)$ has to fulfill the differential equation
\begin{multline}
\left[ \diff[2]{}{r}+ \left( \dfrac{1-2i \omega r^2}{r(r-1)} \right) \diff{}{r} \right. \\ 
\left.- \left(\dfrac{ l(l+1) + (1-s^2)/r}{r(r-1)} \right) \right] R_{\omega s l}(r) = 0 \, . \label{eq:RadialEF}
\end{multline}
This radial equation exhibits two regular singularities which are located at $r=0$ and $r=1$, since these points are poles of first order in the second term and poles of at most second order in the last term. By means of the substitution $r \mapsto 1/ \zeta$ we obtain
\begin{multline}
\left[ \diff[2]{}{\zeta} + \left( \dfrac{3\zeta^2-2\zeta-2i \omega}{\zeta^2(\zeta-1)} \right) \diff{}{\zeta} \right. \\
\left. + \left( \dfrac{ l(l+1) +\zeta(1-s^2)}{\zeta^2(\zeta-1) } \right) \right] R_{\omega s l}(\zeta) = 0
\end{multline}
which yields a pole of second order at $\zeta = 0$ in the coefficient of the first derivative term. Hence, we conclude that the original differential equation \eqref{eq:RadialEF} possesses an irregular singularity at $r = \infty$. With two regular singularities at $r=0,1$ and the irregular singularity at $r=\infty$ the radial equation in Eddington-Finkelstein coordinates belongs to the class of singly confluent Heun equations as well.
It is indeed possible to find a substitution that transforms \eqref{eq:RadialEF} into a standard form of the CHE. One such possibility simply is
\begin{align}
R_{\omega s l}(r) = r^{s+1} y_{\omega s l}(r)  \label{eq:Ansatz_EF}
\end{align}
and yields a CHE in the form \eqref{eq:CHE} with the set of parameter
\begin{align}
\alpha = (s+1) \, , \quad \beta &= 2i\omega \, , \quad \gamma = 1+2s \, , \notag \\
\delta = 1-2i\omega \, , \quad q &= s(s+1) - l(l+1) \, . \label{eq:EF_Parameter_CHE}
\end{align}
\subsection{Local solutions}
\begin{figure}[h!]
\begin{center}
\includegraphics[width=0.4\textwidth]{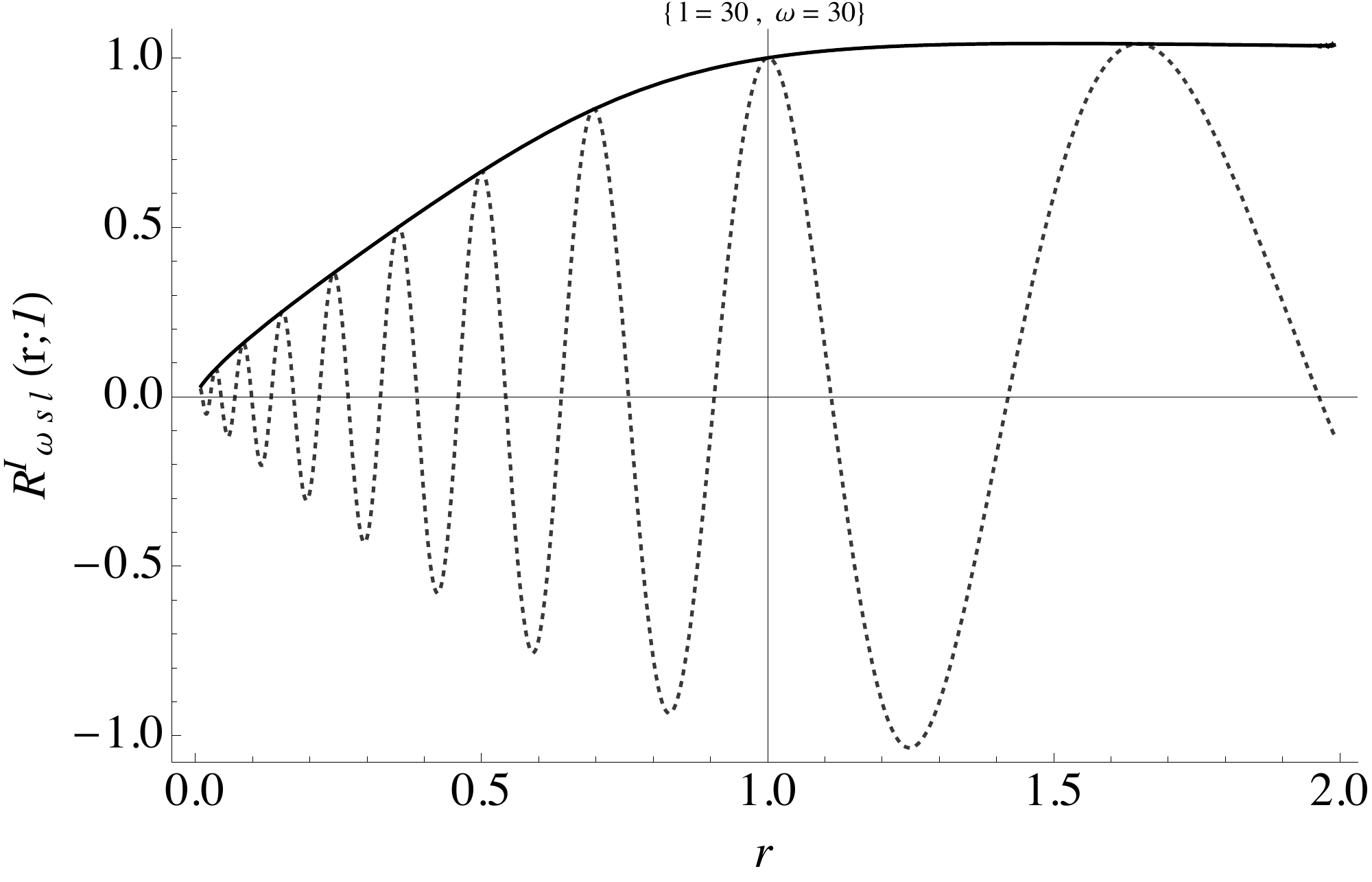}
\includegraphics[width=0.4\textwidth]{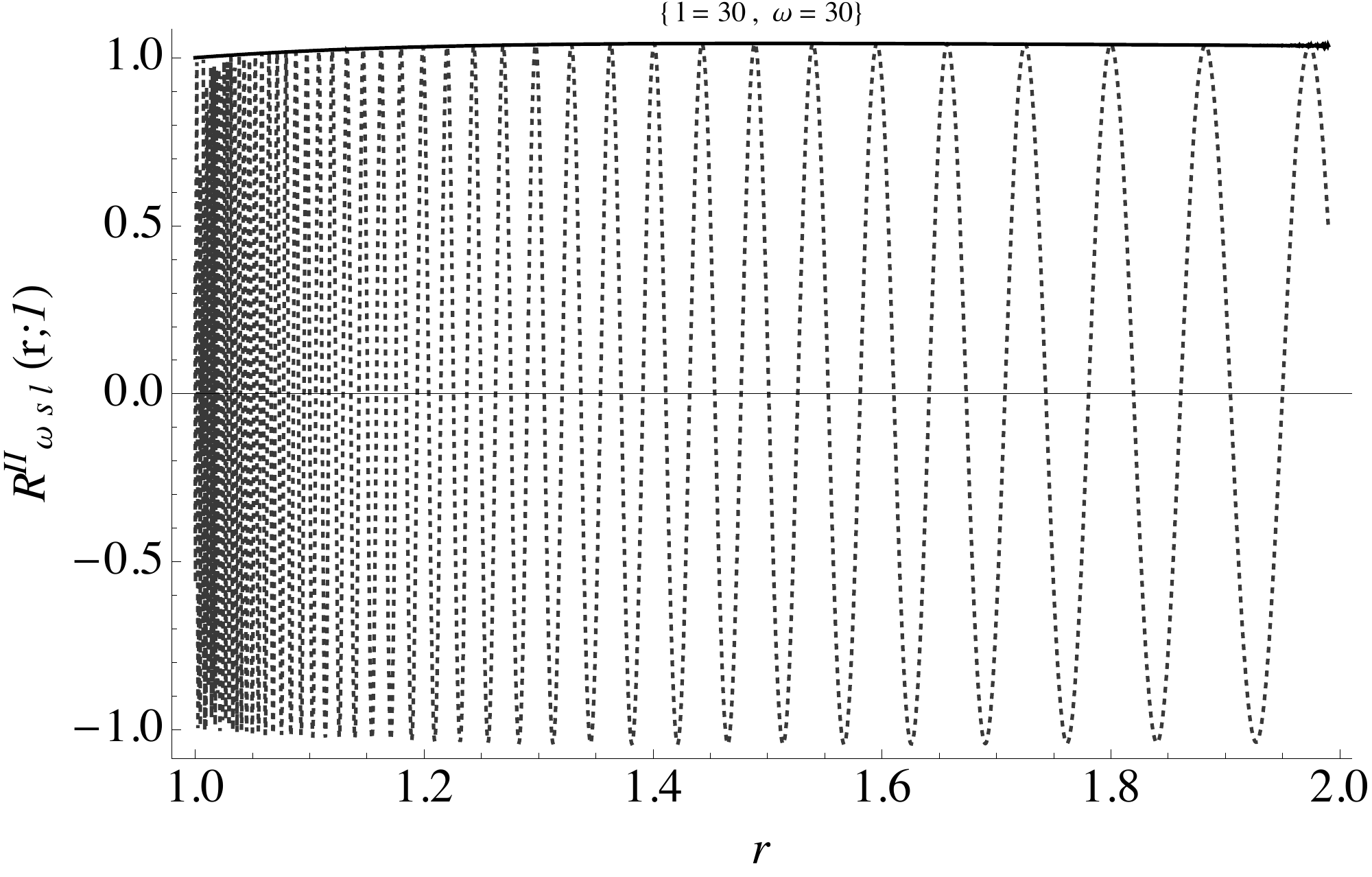}
\end{center}
\caption{\label{fig:EF_Frobenius} The two local solutions around the horizon for $\omega = 30, l=30, s=0$: the modulus (solid) and the real part (dotted) are shown.}
\end{figure}

Using the five parameters in \eqref{eq:EF_Parameter_CHE} we can immediately give the corresponding GRS for the differential equation following the general result \eqref{eq:CHE_GRS},
\begin{align}
\begin{pmatrix}
1 & 1 & 2 &  \\ 
0 & 1 & \infty & ;r \\ 
0 & 0 & s+1 & \\ 
-2s & 2i\omega & s+1-2i\omega&  \\ 
 &  & 0 &  \\ 
 &  & 2i\omega & 
\end{pmatrix} \, .\label{eq:GRS_EF}
\end{align}
This GRS already reveals a lot of useful information: Since $s$ can only take one of the integer values $0,1,2$ the two indicial exponents at the regular singularity $r=0$ differ only by an integer. Hence, both local Frobenius solutions constructed at this point are not linearly independent anymore \cite{MapleHeun}.
Moreover, we can already see that it will be possible to construct a solution that is regular at the horizon. This is due to the indicial exponent $0$ in the second column of the GRS and the transformation \eqref{eq:Ansatz_EF} that is regular at the horizon $r=1$, unlike in the Schwarzschild coordinate case before. It is also possible to give the radial equation in EF coordinates in the Maple form of the CHE. In this case the parameter for the confluent Heun equation in the form \eqref{eq:CHE_MapleForm} become
\begin{align}
a= -2i \omega, \, b =2s, \, c = -2i \omega, \notag \\
d = 2\omega^2, \, e =s^2 -l (l+1) \, .
\end{align}
Using these five parameters we can give the solutions around the regular singularity at the horizon in terms of the standard confluent Heun function as follows:
\begin{widetext}
\begin{subequations}
\begin{align}
R_{\omega s l}^{I}(r;1) &= r^{s+1} \, y_{\omega s l}^{I}(r;1) = r^{s+1} \, \mathrm{HeunC}(2i\omega,-2i\omega,2s,-2\omega^2,s^2-l(l+1)+2\omega^2,1-r) \, , \label{eq:EF_Frobenius1} \\ 
R_{\omega s l}^{II}(r;1) &= r^{s+1} \, y_{\omega s l}^{II}(r;1) = r^{s+1} (r-1)^{2i \omega} \, \mathrm{HeunC}(2i\omega,2i\omega,2s,-2\omega^2,s^2-l(l+1)+2\omega^2,1-r) \, . \label{eq:EF_Frobenius2}
\end{align}
\end{subequations}
\end{widetext}
Both solutions are independent and differ remarkably in their behavior at the horizon: While the first solution is well behaved and takes simply a constant value, the second solution $R_{\omega s l}^{II}(r;1)$ performs infinitely many turns on the unit circle in the complex plane and has no well defined phase when approaching $r=1$. Figure (\ref{fig:EF_Frobenius}) shows the real part and the absolute value of both solutions for some arbitrarily chosen values of the parameter $\omega, s, l$.

Around the irregular singularity at spatial infinity we can construct the Thom\'{e} solutions, using the characteristic exponents of second kind in the third column of the GRS (\ref{eq:GRS_EF}). The result is
\begin{subequations}
\begin{alignat}{2}
R^{I}_{\omega s l}(r;\infty) &= \mathrm{e}^{2i \omega (r+\log(r))} & &\sum_{k=0}^\infty \rho_k(\omega,s,l) ~ r^{-k} \, , \\
R^{II}_{\omega s l}(r;\infty) &= & &\sum_{k=0}^\infty \sigma_k(\omega,s,l) ~ r^{-k}
\end{alignat}
\end{subequations}
and the coefficients $\rho_k$ and $\sigma_k$ can be calculated using recurrence relations obtained from the differential equation, again.

We labeled all the solutions such that we get the same scheme (\ref{fig:InOut_Schwarzschild}) as in the Schwarzschild case before:
The regular solution at the horizon $R_{\omega s l}^{I}(r;1)$ describes waves that propagate into the horizon, while the other solution $R_{\omega s l}^{II}(r;1)$ is associated with waves that emerge from the horizon and propagate to larger radii.
At the irregular singularity at spatial infinity the two solutions $R^{I,II}_{\omega s l}(r;\infty)$ describe in- and outgoing waves as well: The first solution $R^{I}_{\omega s l}(r;\infty)$ describes waves that propagate into the singularity (outgoing) and the second solution $R^{II}_{\omega s l}(r;\infty)$ is associated with waves that emerge from spatial infinity and propagate to smaller radii (ingoing).
\subsection{Analytic continuation of the local solutions}
It is quite unpleasant that the two local solutions $R^{I,II}_{\omega s l}(r;1)$ in (\ref{eq:EF_Frobenius1}) and (\ref{eq:EF_Frobenius2}) have such a small radius of convergence which makes them useful only in the region $|1 - r | < 1$. This is a consequence of the fact that the only known representation of the confluent Heun function is the one in terms of a Frobenius series. When we consider scattering of radiation at a black hole or when we are interested in its quasi-normal modes we have to impose boundary conditions at spatial infinity, or at least at a sufficiently large radius, and therefore the representation of the confluent Heun function $\mathrm{HeunC}$ as a Frobenius series is not appropriate.

In this section we will work out a procedure that is related to a similar problem considered by Jaff\'{e} \cite{Jaffe1934}. He calculated the bound states and energy eigenvalues of the hydrogen molecule ion. Actually, the same idea was applied by Leaver \cite{Leaver1985,Leaver1986} to obtain a continued fraction expression for quasi-normal modes of black holes. Lay \cite{Lay1994} and Slavyanov and Lay \cite{SlavyanovLayBook} gave a general guide for applying the method of Jaff\'{e} to any kind of Heun equation, including all the confluent cases, and they worked out what is called the central two point connection problem (CTCP) in great detail.
The derivations in this section are based on the work of Slavyanov and Lay and we will put our radial equation in such a form that the results in \onlinecite{SlavyanovLayBook} can be used.

The question of interest is now: is it possible to ``connect'' the local solutions at the horizon and those at spatial infinity and obtain expressions that are valid on a larger domain or cover even the entire positive real axis as the domain of convergence? The basic idea is to ``glue'' Frobenius- and Thom\'{e}-type solutions together. In this way we encounter a special case of the central two point connection problem.
To solve this problem we will apply the following five step-procedure:
\begin{enumerate}
\item Shift the regular singularity and reorder the differential equation.
\item Perform an s-homotopic transformation of the dependent variable that involves the indicial exponent of the respective Frobenius solution at the horizon and the characteristic Thom\'{e} exponents (of order zero and one) at spatial infinity.
\item M\"obius-transform the independent variable such that the regular singularity remains unchanged at the origin, but the irregular singularity is moved to the boundary of the unit circle.
\item Solve the resulting differential equation using a series expansion around the origin that converges on the open unit circle.
\item Construct, thereupon, the total solution of the original radial equation by ``inverting'' the previous transformations.
\end{enumerate}
\subsubsection{Shift the singularities and reorder the differential equation}
We use the transformation $r \mapsto z = r - 1$ that maps the regular singularity from $r=1$ to $z=0$, while the irregular singularity remains at $z=\infty$. Here it is important that no other singularity is located between those two on the positive real axis. This yields the differential equation
\begin{multline}
\left[ \diff[2]{}{z} + \left( \dfrac{A_2}{z+z_*} + \dfrac{A_1}{z} + G_0 \right) \diff{}{z} \right. \\
\left. + \left( \dfrac{C_2}{z+z_*} + \dfrac{C_1}{z} \right) \right] y_{\omega s l}(z) = 0
\end{multline} 
with the parameters
\begin{align}
z_* &= 1 \, , \, G_0 = -2i \omega \, , \, A_1 = 1-2i\omega \, , \, A_2 = 1+2s \, , \notag \\
C_1 &= s(s+1) - l(l+1) -2i \omega(s+1) \, ,  \\
C_2 &= l(l+1) - s(s+1) \, , \notag
\end{align}
and is done to match the conventions used in \cite{SlavyanovLayBook}.
\subsubsection{S-homotopic transformation of the dependent variable} 
\begin{figure}
\begin{center}
\includegraphics[width=0.45\textwidth]{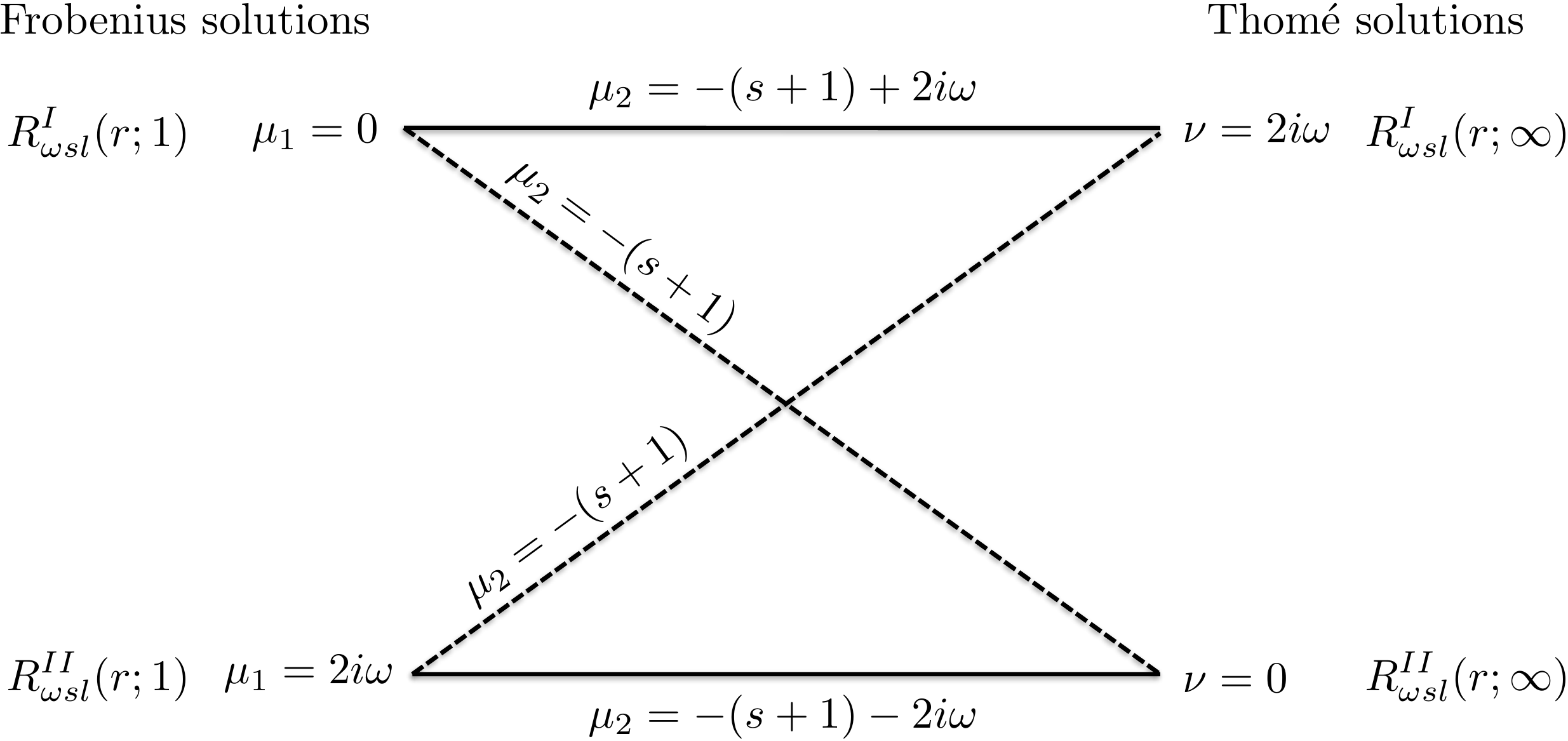}
\end{center}
\caption{\label{fig:exponents} A scheme showing the proper choice of the exponents $\nu, \mu_1$ and $\mu_2$ related to local solutions for the connection problem.}
\end{figure}

We perform an s-homotopic transformation of the dependent variable $y_{\omega s l}(z)$ that contains indicial exponents at $z=0$ and characteristic exponents of the second kind (order zero and order one) at $z=\infty$. This transformation is given by
\begin{align}
&y_{\omega s l}(z) \mapsto \Xi_{\omega s l}(z) \notag \\
&y_{\omega s l}(z) = \mathrm{e}^{\nu z} \, z^{\mu_1} \, (z+1)^{\mu_2} \, \Xi_{\omega s l}(z) \, ,
\end{align}
where the exponents $\nu, \mu_1, \mu_2$ have the following meaning: 
\begin{itemize}
\setlength{\parskip}{1pt}
\item[-] $\mu_1$ is the indicial exponent of the chosen Frobenius solution at $z = 0$. Thus, we can have either $\mu_1 = 0$ or $\mu_1 = 2i \omega$ according to the GRS \eqref{eq:GRS_EF}.
\item[-] $\nu$ is the characteristic Thom\'{e} exponent of order one and depends on the chosen Thom\'{e} solution at $z=\infty$. Hence, we can have either $\nu = 0$ or $\nu = 2i \omega$.
\item[-] For large $z$ we get $z^{\mu_1} (z+1)^{\mu_2} \approx z^{\mu_1+\mu_2}$ and we have to adjust the sum $\mu_1 + \mu_2$ to be the characteristic exponent of order zero of the chosen Thom\'{e} solution. Hence, the possible values for $\mu_2$ are $\mu_2 = -(s+1)$ or $\mu_2 = -(s+1)\pm 2i \omega$ .
\end{itemize}
The scheme in Fig.  (\ref{fig:exponents}) shows which combination of the three exponents must be taken for the desired choice of Frobenius and Thom\'{e} solutions to be connected.
\subsubsection{M\"obius transformation of the independent variable} 
Next, we apply the M\"obius transformation 
\begin{align}
z \mapsto x = \dfrac{z}{z+1}
\end{align}
that maps the irregular singularity from $z=\infty$ to $x=1$ and leaves the regular singularity at $x=0$. This yields a new differential equation for $\Xi(x)$, that is
\begin{multline}
x^2(x-1)^2 \, \diff[2]{\Xi_{\omega s l}(x)}{x} + \sum_{k=1}^3 \Omega_k x^k \, \diff{\Xi_{\omega s l}(x)}{x} \\
+ \sum_{k=1}^2 \Delta_k x^k \, \Xi_{\omega s l}(x) = 0 \, ,
\end{multline}
where the coefficients $\Omega_k$ and $\Delta_k$ are given by
\begin{align}
\Omega_1 &= 1 - 2i\omega + 2\mu_1 \, , \notag \\
\Omega_2 &= 2(s + \nu + \mu_2 - \mu_1 -1) \, , \notag \\
\Omega_3 &= 1 - 2s - 2\mu_2 \, , \notag \\
\Delta_1 &= s(s+1) - l(l+1) + 2(s + \mu_2)(\mu_1-\nu) \notag \\
&-\nu + \mu_1 + \mu_2 \,  \notag \\
\Delta_2 &= \mu_2(2s + \mu_2) \, .
\end{align}
\subsubsection{A series expansion ansatz} 
We use a series expansion ansatz for the unknown function $\Xi_{\omega s l}(x)$
\begin{align}
\Xi_{\omega s l}(x) = \sum_{k=0}^\infty \xi_k(\omega,s,l) ~ x^k \label{eq:CTCP_ansatz}
\end{align}
and insert this ansatz into the differential equation. Thus, we obtain a three term recurrence relation for the series coefficients $\xi_k$
\begin{align}
\xi_0 &= \text{arbitrary} \notag \\
\alpha_0 \xi_1 + \beta_0 \xi_0 &= 0 \notag \\
\alpha_k \xi_{k+1} + \beta_k \xi_k + \gamma_k \xi_{k-1} &= 0 \, , \label{eq:CTCP_recursion}
\end{align}
where $\forall k \geq 0$ we have
\begin{align}
\alpha_0 &= \Omega_1+1 \, , \quad \beta_0 = \Delta_1 \, , \notag \\
\alpha_k &= 1 + \dfrac{\Omega_1+1}{k} + \dfrac{\Omega_1}{k^2} \, , \notag \\
\beta_k &= -2 + \dfrac{\Omega_2+2}{k} + \dfrac{\Delta_1}{k^2} \, , \notag \\
\gamma_k &= 1 - \dfrac{\Omega_3-3}{k} + \dfrac{\Delta_2-\Omega_3+2}{k^2} \, .
\end{align}
The solution (\ref{eq:CTCP_ansatz}) with the recurrence relation (\ref{eq:CTCP_recursion}) is absolutely convergent on the open unit circle around $x=0$, i.e., the radius of convergence is the distance between the two considered singularities at $x=0$ and $x=1$. Choose any $x$ with $|x| < 1$, then
\begin{align*}
\lim_{k\to\infty} \left| \dfrac{\xi_{k+1} x^{k+1}}{\xi_k x^k} \right| = |x| < 1
\end{align*}
and convergence for $|x| < 1$ is assured \cite{Leaver1986}. However, uniform convergence will strongly depend on the properties of the coefficients $\xi_k(\omega,s,l)$, since
\begin{align*}
\lim_{k\to\infty} \left[ \lim_{x\to 1} \left| \dfrac{\xi_{k+1} x^{k+1}}{\xi_k x^k} \right| \right] = \lim_{k\to \infty} \left| \dfrac{\xi_{k+1}}{\xi_k} \right] \, .
\end{align*}
Hence, the solution is uniformly convergent if $\sum_{k} \xi_k(\omega,s,l)$ exists and is finite, which will depend on the values of $\omega,s$ and $l$. The sum will be finite and, thus, uniform convergence will be assured if $\xi_k(\omega,s,l)$ is a minimal solution of the recurrence relation (\ref{eq:CTCP_recursion}) \cite[see, e.g.,][]{SlavyanovLayBook,Leaver1986}.
\subsubsection{Final solution of the radial equation} 
\begin{figure}[h!]
\begin{center}
\includegraphics[width=0.4\textwidth]{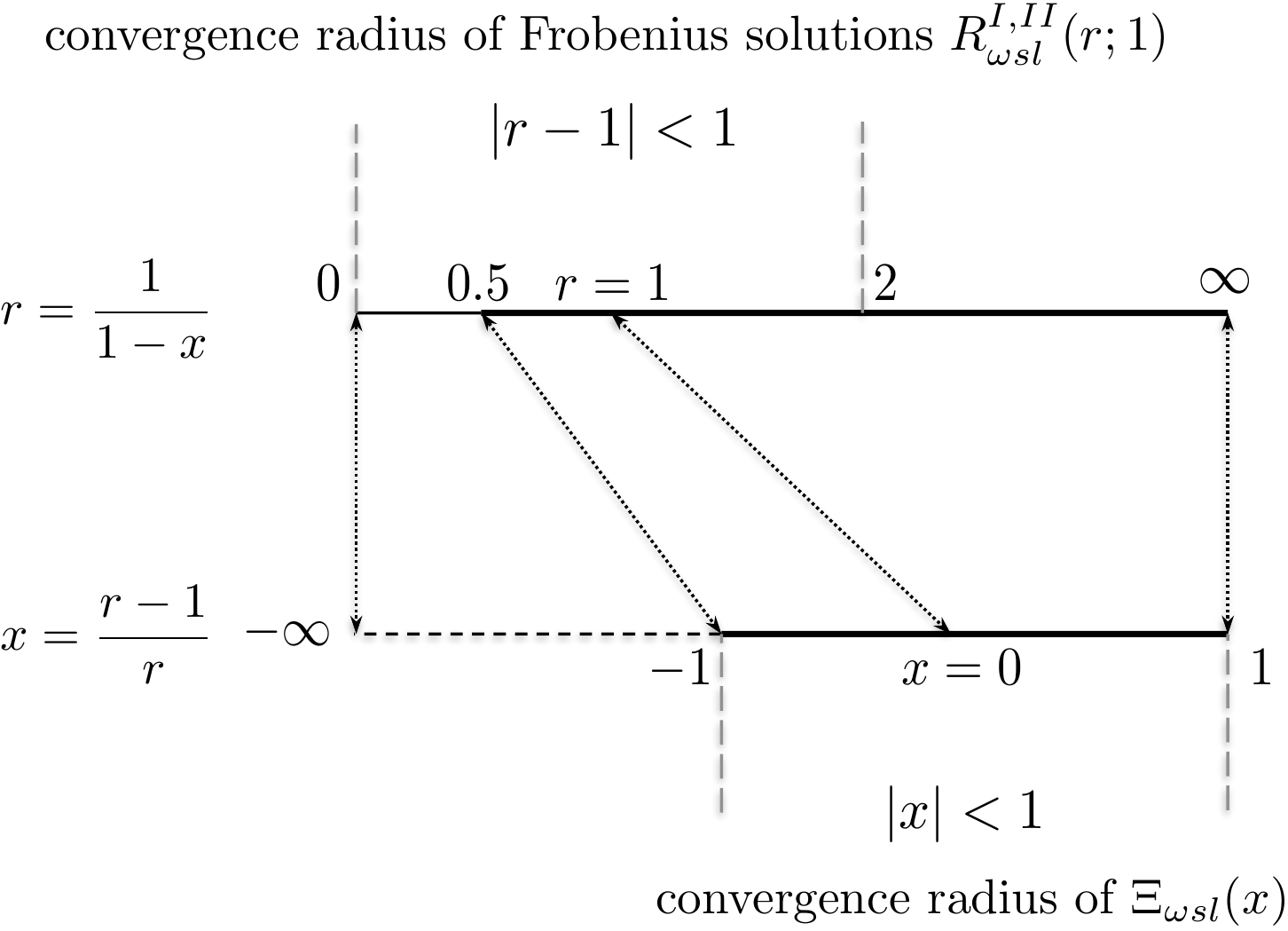}
\end{center}
\caption{\label{fig:mapping} A scheme showing the mapping from $x$ to $r$ and the solutions $R_{\omega s l}(r)$ and $\Xi_{\omega s l}(x)$ together with their domains of convergence.}
\end{figure}

Finally, we can construct the total solution to the radial equation in EF coordinates (\ref{eq:RadialEF}).  We have to use 
\begin{align}
x = \dfrac{z}{z+1} = \dfrac{r-1}{r}
\end{align}
to recover the original radius variable $r$. Thereupon, we obtain
\begin{align}
R_{\omega s l}(r) = \mathrm{e}^{\nu (r-1)} (r-1)^{\mu_1} r^{\mu_2+s+1} \sum_{k=0}^\infty \xi_k(\omega,s,l) \dfrac{(r-1)^k}{r^k} \, .\label{eq:CTCP_solution}
\end{align}
The exponents $\nu, \mu_1,\mu_2$ must be chosen according to the scheme in Fig.~(\ref{fig:exponents}) and the requirements of the physical situation that is to be described. The solution (\ref{eq:CTCP_solution}) is absolutely convergent on the open interval $r \, \in \, ]0.5,\infty[$ owing to the absolute convergence of $\Xi(x)$ for $|x| <1$ and the mapping $x \to r$ as explained in Fig. (\ref{fig:mapping}).
In the entire region where the Frobenius solutions and the analytically continued solutions overlap, they do coincide. Hence, we can interpret our solutions (\ref{eq:CTCP_solution}) as new representations of the confluent Heun function $\mathrm{HeunC}$ with larger domain of convergence. In the next section we show how to use these solutions for different physical applications.
\section{\label{sec:applications} Applications}
\subsection{Black hole scattering}
As an example, we consider the scattering of a massless scalar field at a Schwarzschild black hole. Numerical studies of such scattering processes can be found for example in \cite{KanaiNambu2013} and \cite{Crispino2009}.
In analogy to the usual scattering theory used in quantum mechanics and nuclear physics we consider an ingoing plane wave coming from spatial infinity as initial condition. Such a plane scalar wave is in Schwarzschild coordinates for $r \to \infty$ given by \citep{Andersson1995}
\begin{align}
\Phi_{\text{plane}} \sim \mathrm{e}^{-i\omega t} \sum_{l=0}^\infty \dfrac{i^l (2l+1)}{\omega r}  \sin \left(\omega r_* - \frac{l\pi}{2} \right) P_l(\cos \vartheta) \, .
\end{align}
This is a superposition of partial waves with angular momentum index $l$. Each partial wave will be reflected and/or absorbed by the black hole in a different manner and this will change the composition of the total wave field since different modes will be modified with different amplitudes that refer to scattering and absorption coefficients. This can be best understood by regarding the RWE (\ref{eq:RWE}) and the involved potential. For a given frequency $\omega$ of the incoming wave the potential depends on the wave index $l$ as shown in Fig.~(\ref{fig:RWEpotential2}). As long as the energy $\omega^2$ of the incoming wave is larger than the potential peak for a given $l$-mode, this mode will be (mostly) absorbed by the black hole. When the energy becomes smaller than the potential peak for a fixed $l$-mode, this mode will be (mostly) reflected by the potential barrier. For the special case $\omega^2 = V_{l}^{max}$ we expect the transmission and reflection probabilities to be $\mathbb{T}_{\omega l} \approx \mathbb{R}_{\omega l} \approx 0.5$.

\begin{figure}
\begin{center}
\includegraphics[width=0.4\textwidth]{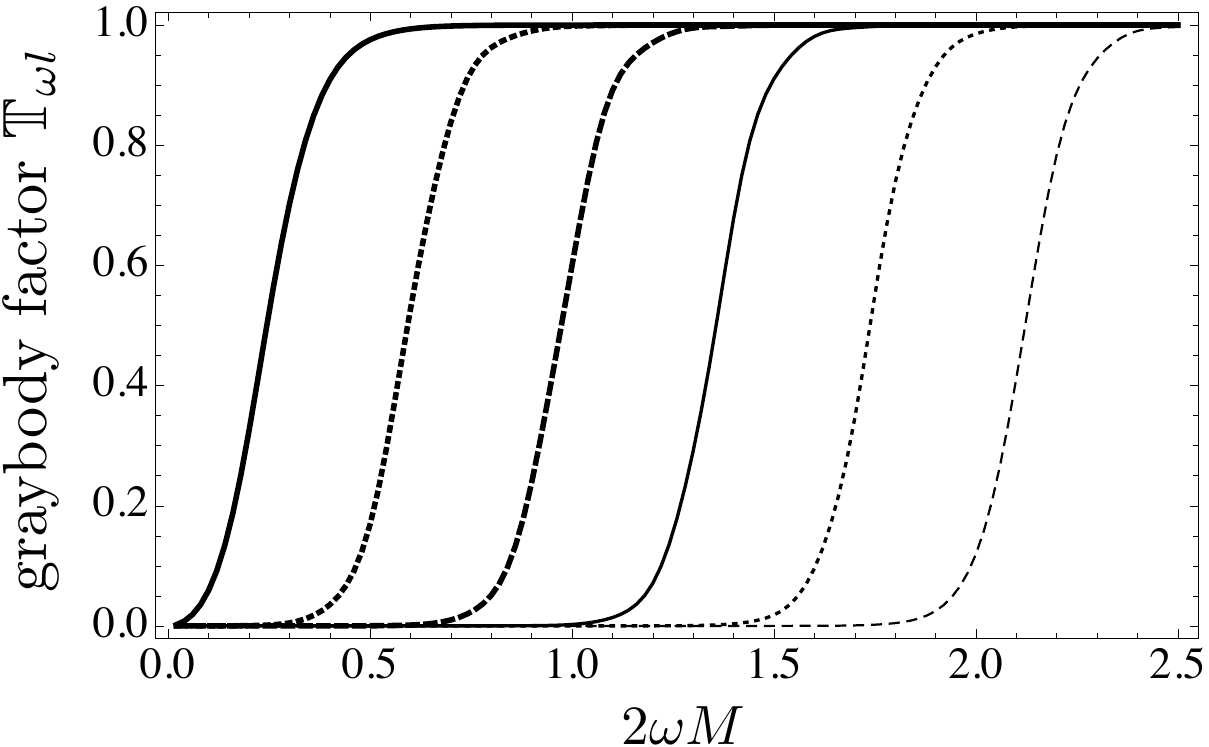}
\end{center}
\caption{\label{fig:graybody} Graybody factors as functions of the frequency for $l=0$ (thick solid) up to $l=5$ (thin dashed).}
\end{figure}

\begin{figure}
\begin{center}
\includegraphics[width=0.4\textwidth]{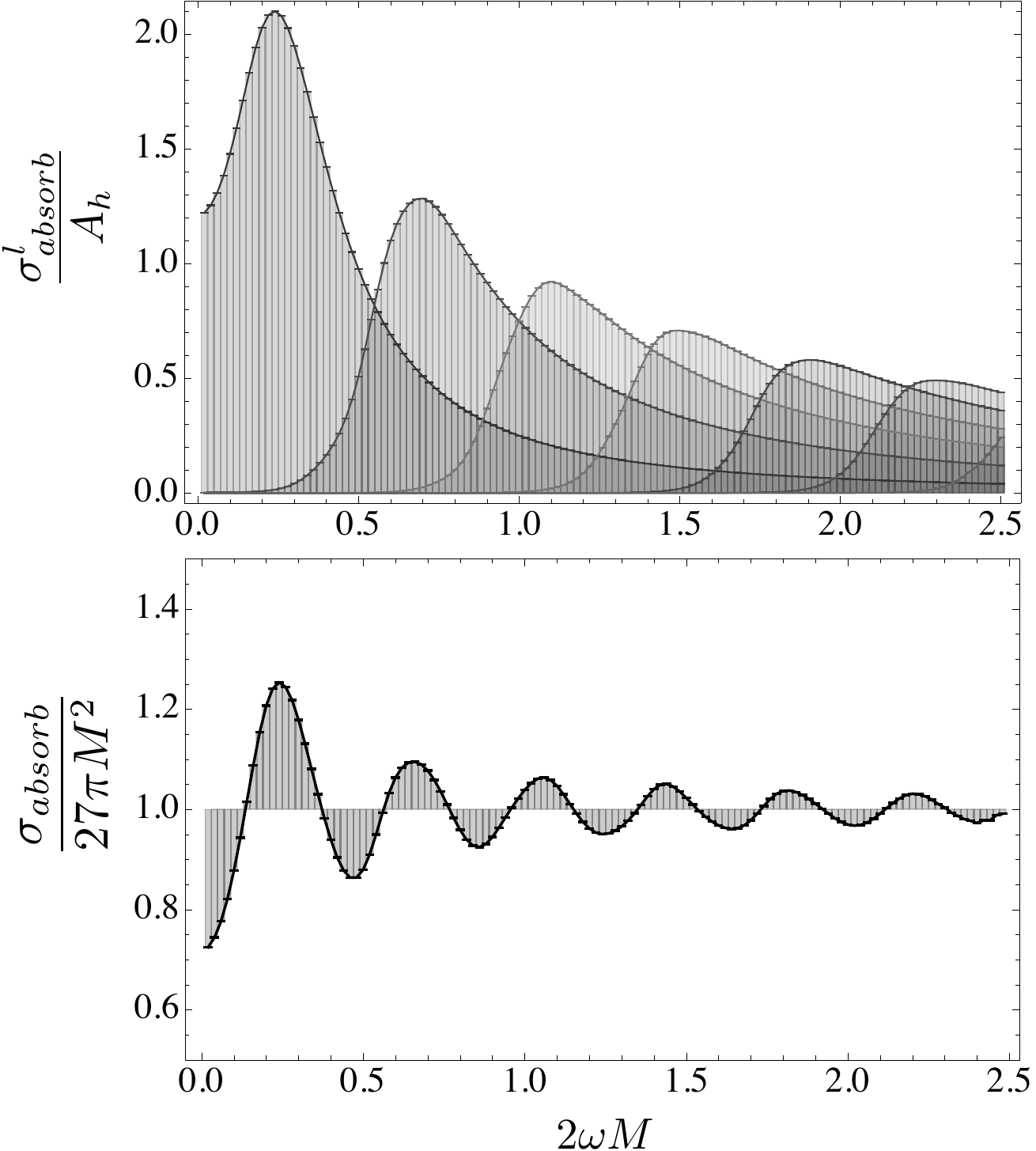}
\end{center}
\caption{\label{fig:partialAbsorption} Top: partial absorption cross sections normalized by the horizon area $A_h$ for wave modes starting from $l=0$ (left) up to $l=5$; Bottom: total absorption cross section as a sum over all contributions normalized by the geometrical optics value $27 \pi M^2$.}
\end{figure}

To derive the transmission and absorption probabilities we must use the solution (\ref{eq:CTCP_solution}) with the exponents chosen to be
\begin{multline}
\nu = 0 \, , \quad \mu_1 = 0 \, , \quad \mu_2 =-(s+1) \\
\Rightarrow R_{\omega s  l}(r) = \sum_{k=0}^\infty \xi_k(\omega,s,l) \dfrac{(r-1)^k}{r^k} \label{eq:Scattering_radialFunction}
\end{multline}
to match the causal boundary condition, i.e., to ensure purely ingoing waves at the horizon. The full wave field in EF coordinates is then given by
\begin{align}
\Phi \sim \mathrm{e}^{-i\omega v} \sum_{l=0}^\infty \dfrac{(2l+1)}{r} R_{\omega s l}(r) P_l(\cos \vartheta) \, .
\end{align}
We can, of course, transform this solution to usual Schwarzschild coordinates simply by using $v=t+r_* = t+r+\log(r-1)$. This might be done since the plane wave is given in Schwarzschild coordinates as well. (The other option is to give the plane wave in EF coordinates.) Thus, we have
\begin{align}
\Phi \sim \mathrm{e}^{-i\omega t} \mathrm{e}^{-i\omega (r+\log(r-1))} \sum_{l=0}^\infty \dfrac{(2l+1)}{r} R_{\omega s l}(r) P_l(\cos \vartheta) \label{eq:Scattering_WaveField}
\end{align}
and we know that in the limit $r \to \infty$ the asymptotic form must be
\begin{align}
\Phi \sim \mathrm{e}^{-i\omega t} \sum_{l=0}^\infty \dfrac{(2l+1)}{r} \left(A^{in}_{\omega l} \, \mathrm{e}^{-i \omega r_*} + A^{out}_{\omega l} \, \mathrm{e}^{i \omega r_*} \right) P_l(\cos \vartheta) \, . \label{eq:Scattering_WaveField_asympt}
\end{align}
Now, we can proceed as in usual quantum mechanical scattering theory: At large distances the total wave field shall be the sum of the incoming plane wave and a back-scattered part due to reflection at the black hole
\begin{align}
\Phi \sim \Phi_{\text{plane}} + \mathrm{e}^{-i\omega t} f(\vartheta) \dfrac{\mathrm{e}^{i\omega r_*}}{r} \quad \text{for} \quad r \to \infty \, ,
\end{align}
and we define the complex valued phase shifts $\delta_l$ by
\begin{align}
\Phi - \Phi_{\text{plane}} \sim \mathrm{e}^{-i \omega t} \mathrm{e}^{i\omega r_*} \sum_{l=0}^\infty \dfrac{(2l+1)}{2i\omega r} \left( \mathrm{e}^{2i \delta_l} - 1 \right) P_l(\cos \vartheta) \, .
\end{align}
We can show that this implies
\begin{align}
\mathrm{e}^{2i\delta_l} = (-1)^{l+1} \dfrac{A^{out}_{\omega l}}{A^{in}_{\omega l}} \, .
\end{align}
Hence, we have to calculate the transmission and reflection amplitudes $A^{in, out}_{\omega l}$ to obtain the phase shifts. We can do so by matching our solution (\ref{eq:Scattering_WaveField}) with the radial function given by \eqref{eq:Scattering_radialFunction} to the asymptotic form (\ref{eq:Scattering_WaveField_asympt}). Even though the radial function in the form (\ref{eq:Scattering_radialFunction}) is not in general convergent in the limit $r \to \infty$ for an arbitrary choice of the parameter $\omega,s$ and $l$ we can fit it to the asymptotic behavior on an interval for large values of $r$ and for fixed mode number $l$ and frequency $\omega$. With this procedure we can calculate the phase shifts in a semi-analytical way for all desired combinations of the parameters. Using the phase shifts or the reflection and transmission amplitudes we can calculate the transmission and reflection coefficients
\begin{align*}
\mathbb{T}_{\omega l} = |T|^2_{\omega l} = \left| \dfrac{1}{A^{in}_{\omega l}} \right|^2, \quad \mathbb{R}_{\omega l} = |R|^2_{\omega l} = \left| \dfrac{A^{out}_{\omega l}}{A^{in}_{\omega l}} \right|^2 \, .
\end{align*}
They describe the probability of being transmitted through the potential barrier, i.e., being absorbed by the black hole and the probability of being reflected back to spatial infinity. The transmission coefficients $\mathbb{T}_{\omega l}$ are usually called graybody factors, since they are important when we derive the deviations from a thermal spectrum of Hawking radiation. Figure (\ref{fig:graybody}) shows the calculated graybody factors as functions of the frequency for different wave modes.
Using the graybody factors we can, for example, calculate the total absorption cross section as a sum over all partial contributions
\begin{align}
\sigma_{abs} = \sum \sigma^l_{abs} = \dfrac{\pi}{\omega^2} \sum (2l+1) \mathbb{T}_{\omega l} \, .
\end{align}
In Fig.~(\ref{fig:partialAbsorption}) we show the partial absorption cross sections for different values of the wave mode starting from $l=0$ (on the left) to $l=5$. The results are normalized by the horizon area. The figure further shows the total absorption cross section as a sum over all partial contributions. The result is normalized by the geometrical optics value of the cross section $27\pi M^2$, which is recovered in the limit of high frequencies as it should be. For small frequencies $\omega M \to 0$ the partial absorption cross section for the $l=s=0$ mode (lowest mode) approaches the horizon area of the black hole and therefore overcomes the problem shown in Figure (7) in the phase-integral study by Andersson \cite{Andersson1995}. For higher frequencies our results match those given by Andersson and our results in Fig.~(\ref{fig:partialAbsorption}) are in agreement with those shown in Figures (2) and (3) in the early work by Sanchez \cite{Sanchez1978}.
\subsection{Quasi-normal modes}
We can use the radial solutions (\ref{eq:CTCP_solution}) to obtain the quasi-normal mode frequencies of a Schwarzschild black hole for a spin $s$ perturbation. To ensure the proper QNM boundary conditions we have to choose, according to the scheme in Fig.~(\ref{fig:exponents}), the exponents to be
\begin{multline}
\nu = 2i \omega, \quad \mu_1 = 0, \quad \mu_2 = -(s+1) + 2i \omega \\
\Rightarrow R^{QNM}_{\omega s  l}(r) = \mathrm{e}^{2i \omega (r-1+log(r))}  \sum_{k=0}^\infty \xi_k(\omega,s,l) \dfrac{(r-1)^k}{r^k}
\end{multline}
and we further demand the solution to converge in the limit $r\to \infty$. The convergence property is fulfilled if the series coefficients $\xi_k$ are minimal solutions of the recurrence relation \eqref{eq:CTCP_recursion}. This immediately yields the continued fraction equation that was given by Leaver \cite{Leaver1985} and the QNM frequencies $\omega_{QNM}$ can be calculated as complex roots of this equation. Hence, we see that the method of Leaver fits perfectly into our framework of continued local Frobenius solutions. For details on quasi-normal mode frequencies the reader is referred to Refs. \cite{Berti2006,Cardoso2004} and \cite{Berti2009} and to the living reviews \cite{KokkotasSchmidt,SasakiTagoshi}.
\subsection{Hawking radiation}
Using the method of Damour and Ruffini \cite{DamourRuffini1976} we can calculate the spectrum of the Hawking radiation emitted by a Schwarzschild black hole. Now, we have to consider that part of the solution to the radial equation which was always neglected up to now: We need to use the second solution $R^{II}(r;1)$ that is not regular at the horizon. This solution is given by (\ref{eq:EF_Frobenius2}) and describes purely outgoing waves at the horizon $r=1$: This is the outgoing Hawking radiation that is obtained here as the solution of a classical wave equation on curved background geometry.
Close to the horizon this solution behaves like
\begin{align}
R^{II}(r;1) \sim (r-1)^{2i\omega} \, .
\end{align}
Following Damour and Ruffini, there exists a unique continuation of this solution for radii $r<1$, which yields
\begin{align}
\overline{R} ^{II}(r;1) =
\begin{cases}
(r-1)^{2i\omega} \quad &r>1 \\
\mathrm{e}^{2\pi \omega} (1-r)^{2i\omega} \quad &r<1
\end{cases} \, .
\end{align}
This continuation then describes an antiparticle of negative energy inside going down the black hole and a particle outside emerging from the horizon. The probability for transmission through the horizon is given by the squared modulus of the relative amplitude
\begin{align}
p_{\omega} = \mathrm{e}^{-4\pi\omega} \left( = \mathrm{e}^{-8\pi\omega M} \right) \, .
\end{align}
Using Feynman's method \cite{Feynman1949} of scattering amplitudes one can deduce that this is also the probability for antiparticle-particle pair production just outside $r=1$. In \cite{Sannan1988} Sannan showed how to get the mean number of particles that are radiated to spatial infinity in a given mode from the pair production probability $p_\omega$. The result is
\begin{align}
\overline{N}_{\omega l} = \dfrac{\mathbb{T}_{\omega l}}{\mathrm{e}^{8\pi\omega M} \pm 1} \, ,\label{eq:HawkingRadiationMeanNumber}
\end{align}
where the $+$ is for fermionic particles and the $-$ must be used for bosons. The bosonic result is exactly Hawking's result \cite{Hawking1976} that was obtained using calculations involving the gravitational collapse. When we consider bosons, the spectrum is thermal except for the modification caused by the transmission probability (or graybody factor) $\mathbb{T}_{\omega l}$, which causes an amplitude filtering of different modes. Hence, the total spectrum is not really thermal anymore but is usually called graybody radiation (instead of black-body radiation with thermal spectrum). From (\ref{eq:HawkingRadiationMeanNumber}) we can derive a power spectrum of the emitted radiation. Each massless boson takes away an energy of $\omega$ from the black hole and the change of energy per unit of time and frequency interval is minus the radiated power which is given by
\begin{align}
\dfrac{dE}{d\omega dt} = \dfrac{1}{2\pi} \sum_{l,p} (2l+1)  \dfrac{\mathbb{T}_{\omega l} \, \omega}{\mathrm{e}^{8\pi\omega M} - 1} \, . \label{eq:HawkingRadiation_Spectrum}
\end{align}
Here, $p$ denotes additional degrees of freedom like the polarization of the particle \cite{Sannan1988}. It should be mentioned that this radiation spectrum is valid for a time in which we consider the background geometry to remain unchanged, i.e., the mass of the black hole is taken to be constant during this time interval. Then, we may iterate and consider the mass to be decreased due to the evaporated energy and repeat the calculation of the radiated power with the decreased mass. In this way one can derive the mass of the black hole as a function of time and calculate the life time of black holes with given initial mass.
\begin{figure}
\begin{center}
\vspace{10pt}
\includegraphics[width=0.3\textwidth]{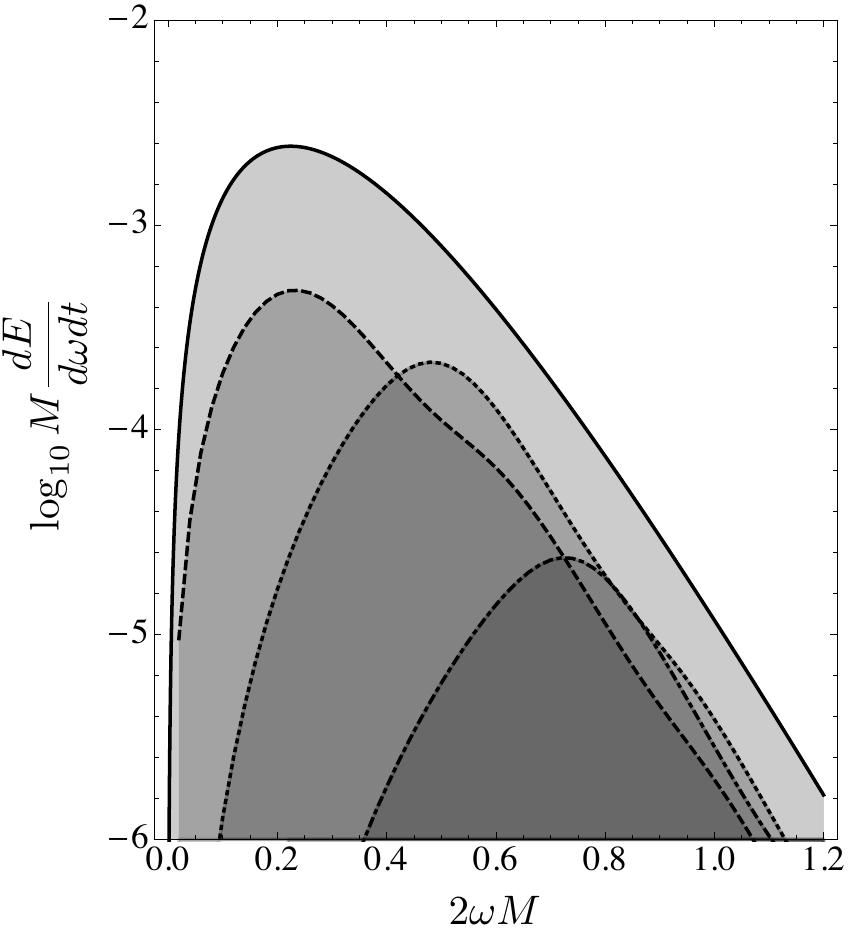}
\end{center}
\caption{\label{fig:hawking} Power spectrum of Hawking radiation emitted by a Schwarzschild black hole for massless bosons with spin $s=0$ (dashed), $s=1$ (dotted), $s=2$ (dash-dotted) compared to a thermal spectrum for a cross section of $27\pi M^2$ (solid).}
\end{figure}
Figure (\ref{fig:hawking}) shows the obtained power spectrum according to (\ref{eq:HawkingRadiation_Spectrum}) for massless bosons with spin $s \in \left\lbrace 0,1,2 \right\rbrace$.
The results in Fig.~(\ref{fig:hawking}) match previous results, for example those obtained by Page \cite{Page1976} using purely numerical calculations. The major contributions in the spectra in Fig.~(\ref{fig:hawking}) are due to the lowest $l=s$ modes for each spin, respectively. The little ``bumps'' that we can see at higher frequencies are due to higher mode contributions. For a fixed spin the maxima of higher mode contributions are shifted to higher frequencies but their peak values decrease rapidly with increasing mode number.
\FloatBarrier
\section*{Conclusions}
The Regge-Wheeler equation is usually written in terms of the tortoise coordinate. However, it can also be written in other coordinate systems, such as usual Schwarzschild, Eddington-Finkelstein, Painlev{\'e}-Gullstrand or Kruskal-Szekeres coordinates. While it is not possible to obtain regular solutions at the black-hole horizon in Schwarzschild coordinates (see the comment following Eq.\eqref{eq:AnsatzFieziev}), we have shown that such solutions can be derived in EF coordinates. This is also possible in PG coordinates where, however,  the expressions are somewhat more involved. These solutions are given in terms of the standard confluent Heun function around the black hole horizon which is defined in terms of Frobenius series. As these series converge on the unit sphere in the complex plane centered at $r=2M$, they cannot be evaluated at larger radii. We have shown how this problem can be overcome by using an analytical continuation procedure.  This gave us analytical solutions in terms of convergent series on the entire domain up to (but not in general including) infinity. Based on these exact solutions we have re-obtained, in a semi-analytic fashion, several results on black-hole scattering, quasi-normal modes and Hawking radiation that had been found numerically earlier. In a follow-up  paper we consider a massive scalar field in EF coordinates which allows us to define purely ingoing solutions that penetrate the horizon. Such a massive scalar field is a possible dark matter candidate and analytical solutions might be of interest in this area. In the same follow-up paper we further re-consider  massless fields of spin $s=0$, 1 or 2 but switch to the PG coordinates which are associated with infalling observers and therefore, at least in principle, with possible measurements.\\
\begin{acknowledgments}
The authors would like to thank Domenico Giulini, Norman G\"urlebeck and Claus L{\"a}mmerzahl for valuable discussions during the preparation of the manuscript. Support from the Deutsche Forschungsgemeinschaft within the Research Training Group 1620 "Models of Gravity" is gratefully aknowledged. The first author is financially supported by the German Collaborative Research Center 1128 ``geo-Q''.
\end{acknowledgments}


\end{document}